\documentclass[prd,twocolumn,showpacs,superscriptaddress,nofootinbib,floatfix,showkeys,10pt]{revtex4-2}
\usepackage{bm}
\usepackage{times}
\usepackage{braket}
\usepackage{amsfonts,amssymb,stmaryrd,latexsym,amsmath}
\usepackage[usenames,dvipsnames]{color}
\usepackage{epsfig}
\usepackage{slashed}
\usepackage{hyperref}
\usepackage{subfigure}
\usepackage{graphicx}
\usepackage{slashed}
\usepackage{orcidlink}
\usepackage{multirow}
\usepackage{appendix}
\usepackage{dashrule}
\usepackage{bbm}
\usepackage[normalem]{ulem}
\usepackage{mdframed}

\allowdisplaybreaks[1]

\definecolor{nicered}{rgb}{0.7,0.1,0.1}
\definecolor{nicegreen}{rgb}{0.1,0.5,0.1}
\definecolor{emph}{rgb}{1,0,0}
\definecolor{doub}{rgb}{0.7,0.2,1.0}
\definecolor{navyblue}{RGB}{0, 110, 184}
\hypersetup{colorlinks,citecolor=nicegreen,linkcolor=nicered,urlcolor=navyblue}

 \newcommand{\clabel}[2][]{#2}
 
\begin{document}

	
	\title{DeepQuark: A Deep-Neural-Network Approach to Multiquark Bound States} 
	\author{Wei-Lin Wu\,\orcidlink{0009-0009-3480-8810}}
	\affiliation{School of Physics, Peking University, Beijing 100871, China}
	\author{Lu Meng\,\orcidlink{0000-0001-9791-7138}}\email{lmeng@seu.edu.cn}	
    \affiliation{School of Physics, Southeast University, Nanjing 211189, China}
	\affiliation{Institut f\"ur Theoretische Physik II, Ruhr-Universit\"at Bochum,  D-44780 Bochum, Germany }
	\author{Shi-Lin Zhu\,\orcidlink{0000-0002-4055-6906}}\email{zhusl@pku.edu.cn}
	\affiliation{School of Physics and Center of High Energy Physics,
		Peking University, Beijing 100871, China}
	
	\begin{abstract}

    For the first time, we implement the deep-neural-network-based variational Monte Carlo approach for the multiquark bound states, whose complexity surpasses that of electron or nucleon systems due to strong SU(3) color interactions. We design a novel and high-efficiency architecture, DeepQuark, to address the unique challenges in multiquark systems such as stronger correlations, extra discrete quantum numbers, and intractable confinement interaction. Our method demonstrates competitive performance with state-of-the-art approaches, including diffusion Monte Carlo and Gaussian expansion method, in the nucleon, doubly heavy tetraquark, and fully heavy tetraquark systems. Notably, it outperforms existing calculations for pentaquarks, exemplified by the triply heavy pentaquark. For the nucleon, we successfully incorporate three-body flux-tube confinement interactions without additional computational costs. In tetraquark systems, we consistently describe hadronic molecule $T_{cc}$ and compact tetraquark $T_{bb}$ with an unbiased form of wave function ansatz. In the pentaquark sector, we obtain weakly bound $\bar D^*\Xi_{cc}^*$ molecule $P_{cc\bar c}(5715)$ with $S=\frac{5}{2}$ and its bottom partner $P_{bb\bar b}(15569)$. They can be viewed as the analogs of the molecular $T_{cc}$. We recommend experimental search of $P_{cc\bar c}(5715)$ in the D-wave $J/\psi \Lambda_c$ channel. DeepQuark holds great promise for extension to larger multiquark systems, overcoming the computational barriers in conventional methods. It also serves as a powerful framework for exploring confining mechanism beyond two-body interactions in multiquark states, which may offer valuable insights into nonperturbative QCD and general many-body physics.
    
	\end{abstract}
	\maketitle

    \emph{Introduction}---The quark model proposed by Gell-Mann~\cite{Gell-Mann:1964ewy} and Zweig~\cite{Zweig:1964ruk} provides a remarkably successful framework for understanding the conventional hadron spectrum, classifying hadrons into mesons ($q\bar q$) and baryons ($qqq$). Although multiquark states such as tetraquarks $(qq\bar q\bar q)$ and pentaquarks ($qqqq\bar q$) were proposed concurrently~\cite{Gell-Mann:1964ewy,Zweig:1964ruk}, their existence remained elusive until recent years. Since the discovery of the $X(3872)$ in 2003~\cite{Belle:2003nnu}, numerous multiquark candidates have been observed, including many manifestly exotic states like $T_{c\bar c 1}(3900)$~\cite{BESIII:2013ris,Belle:2013yex}, $T_{cc}(3875)^+$~\cite{LHCb:2021vvq,LHCb:2021auc} , $P_{c\bar c}$ states~\cite{LHCb:2015yax, LHCb:2019kea}, and $T_{cc\bar c\bar c}(6900)^0$~\cite{LHCb:2020bwg} (see~\cite{Chen:2016qju,Hosaka:2016pey,Lebed:2016hpi,Guo:2017jvc,Liu:2019zoy,Brambilla:2019esw,Chen:2022asf,Meng:2022ozq,Wang:2025sic} for comprehensive reviews).  Elucidating the configurations of these multiquark states poses new challenges to the quark model, and also offers a unique window into the nonperturbative regime of quantum chromodynamics (QCD)—the fundamental theory of strong interactions. The investigation of clustering behaviors in multiquark systems, particularly in distinguishing compact configurations from loosely bound hadronic molecules, offers an unprecedented opportunity to explore quantum many-body dynamics governed by SU(3) color interactions, which is fundamentally different from systems dominated by electromagnetic or nuclear forces.

     Solving the quantum many-body problem of multiquark states in quark models can be quite challenging.  In addition to the exponential scaling of wave function dimensionality with the particle number, the multiquark systems involve an extra SU(3) color degree of freedom compared to electron and nucleon systems, causing even greater complexity. Moreover, constructing multiquark systems of interest often requires imposing constraints from various quantum numbers—such as spin, parity, flavor, and color—that arise from underlying symmetry principles. These constraints significantly complicate the theoretical treatment. For example, in calculating the ground states of electron systems, spins are usually not explicitly constrained, as the ground state naturally converges to a certain spin configuration. However, in the case of  doubly charmed tetraquark $T_{cc}$~\cite{LHCb:2021vvq,LHCb:2021auc}, the total spin should be constrained to 1 to avoid falling into the lower scalar $DD$ threshold. Furthermore, the strong color interaction between (anti)quarks leads to significant correlations, rendering the single-particle approximation—effective in nuclear and atomic systems—invalid. Consequently, the shell structure observed in those systems is absent in multiquark states. Without such a simplified approximation as a  guidance or starting point, the full dynamical multichannel treatments of multiquark systems become  computationally intensive.  Additionally, progress in lattice QCD~\cite{Okiharu:2004ve,Cardoso:2012uka} seems to favor the few-body confinement mechanisms over pairwise interactions in multiquarks, introducing new challenges to their quantum many-body descriptions~\cite{Bicudo:2015bra}. The aforementioned complexities make comprehensive calculations of multiquark states highly challenging, even for five-body systems. 
    Existing approaches, such as basis expansion methods exemplified by the Gaussian expansion method (GEM)~\cite{Hiyama:2003cu}, are hindered from complete calculations due to the exponential growth of basis states. Meanwhile, diffusion Monte Carlo (DMC) suffers from the notorious sign problem~\cite{Troyer:2004ge}, which severely limits its applicability to such strongly correlated systems. Previous studies on pentaquarks~\cite{Hiyama:2018ukv,Giron:2021fnl,Yan:2021glh,Yang:2022bfu,An:2022fvs,Liang:2024met,Gordillo:2024blx} made various approximations in the spatial configurations~\cite{Giron:2021fnl,Yan:2021glh,Yang:2022bfu,An:2022fvs,Liang:2024met} or color degree of freedom~\cite{Hiyama:2018ukv,Giron:2021fnl,Gordillo:2024blx} to simplify the calculations, which may result in unknown systematical errors and unreliable conclusions. 

    Recently, the development of machine learning techniques provides a new approach to solving the quantum many-body problem. Motivated by the exceptional capacity of deep neural networks (DNNs) to approximate high-dimensional functions~\cite{LeCun:2015pmr}, DNNs are anticipated to be a flexible and effective wave function representation capable of capturing complex many-body correlations efficiently~\cite{Carleo:2019ptp,zhang2025AI4S}. DNN-based variational Monte Carlo (VMC) method has been successfully applied to quantum spin systems~\cite{Carleo2017}, atomic and molecular physics~\cite{Han2019,FermiNet,hermann2020deep}, condensed matter~\cite{li2022ab,Kim:2023fwy}, and nuclear physics~\cite{Keeble2020,Adams:2020aax,Yang:2022esu,Yang:2022rlw,Yang:2024wsg,Fore:2024exa,Yang:2025mhg}, achieving high accuracies and demonstrating potential for scaling to larger systems. However, its application in multiquark 
     remains unexplored. \clabel[B1-1]{DNN-based wave functions, which are free from \textit{a priori} assumptions about multiquark configurations, can unbiasedly and consistently describe various multiquark states, including hadronic molecules and compact multiquarks. This capability is supported by the Universal Approximation Theorem of neural networks~\cite{HORNIK1989359,kidger2020} and demonstrated by our architecture and numerical results as follows.} Moreover, VMC circumvents the sign problem in imaginary time evolution of DMC, making it capable of handling strongly correlated systems. Meanwhile, unlike basis expansion methods, VMC enables the treatments of complex interactions that go beyond two-body forces even without extra computational costs, paving the way for investigations of confining mechanisms and many-body forces in multiquark states.

    In this work, we develop a DNN-based VMC approach, DeepQuark, to calculate multiquark bound states in the quark model.  A major difference between DeepQuark and previous DNN-based studies is the architecture of many-body wave function. Most of the previous works use a determinant type wave function ansatz~\cite{FermiNet,hermann2020deep,li2022ab,Yang:2022rlw}, which originates from the idea of single-particle orbitals in electron and nucleon systems, and consider the correlations by multiple determinants, Jastrow factors and backflow transformation. However, such constructions may not apply to the strongly correlated hadron systems. Instead, we construct the DeepQuark wave function in the coupled color-spin-isospin bases, which represents the correlations in the most general way while automatically enforcing symmetry requirements. We first benchmark our results against GEM and DMC for the nucleon in two different confinement interactions and tetraquark bound states. We further demonstrate DeepQuark's ability for larger systems by investigating triply heavy pentaquarks $QQqq\bar Q$ ($Q=b,c;q=u,d$), where bound states that are analogs of the doubly heavy tetraquarks $QQ\bar q\bar q$ are obtained.

	\emph{Hamiltonian}---\clabel[A1-1]{We employ the nonrelativistic Hamiltonian in a minimal quark model—the AL1 model (where ``L" denotes linear confinement interaction) introduced in Refs.~\cite{Semay:1994ht,SilvestreBrac1996}. The AL1 potential consists of two parts: the one-gluon-exchange interaction $V_{\text{OGE}}$, accouting for the short-range perturbative QCD effects, and a two-body linear confinement term $V_{\text{conf}}$, which models the quark confinement in hadrons.}
	\begin{equation}
		\label{eq:AL1}
		\begin{aligned}
			&V_{\text{OGE},ij} =-\frac{3}{16} \boldsymbol\lambda_i \cdot \boldsymbol\lambda_j\left(-\frac{\kappa}{r_{i j}}-\Lambda+\frac{8 \pi \kappa^{\prime}}{3 m_i m_j} \frac{e^{ -r_{i j}^2 / r_0^2}}{\pi^{3 / 2} r_0^3} \boldsymbol{s}_i \cdot \boldsymbol{s}_j\right),\\
			&V_{\text{conf},ij}=-\frac{3}{16} \boldsymbol\lambda_i \cdot \boldsymbol\lambda_j \lambda r_{i j}.
		\end{aligned}
	\end{equation}
	Here, $\boldsymbol\lambda_i $ is the SU(3) Gell-Mann color matrix acting on the $i$ th quark (replaced by -$\boldsymbol\lambda^*$ for antiquark), and $\boldsymbol{s}_i$ is the spin operator. This interaction structure closely resembles that of the well-known Cornell model~\cite{Eichten:1978tg, Eichten:1979ms} but takes the light quark sector into consideration. \clabel[A1-2]{The parameters of the model, including four constituent quark masses and six potential parameters, were determined by fitting the meson spectra across all flavor sectors.}  
	
	For baryon systems, an additional phenomenological term $V_{123}=-\frac{C}{m_1m_2m_3}$ is introduced to mimic the three-body interaction effect~\cite{SilvestreBrac1996}. For hadron systems with heavy quarks, this term is suppressed by the heavy quark mass and therefore is neglected in the calculations. For the nucleon, we also test the flux-tube confinement interaction $V^{\text{ft}}_{\text{conf}}=\sigma L_{\text{min}}$, where $L_{\text{min}}$ is the minimal length of the color flux tubes connecting three quarks at a junction point~\cite{Artru:1974zn,Takahashi:2000te,Takahashi:2002bw}. More details on the interactions, the model parameters, and the model uncertainty are given in Supplemental Material~\cite{Suppl}.
	
	In the absence of spin-orbit and tensor operators, the total angular momentum $J$, total orbital angular momentum $L$, and total spin $S$ are all conserved. Since ground states are expected to be S wave, we have $J=S$.

	\begin{figure*}[tbp]
		\centering
		\includegraphics[width=0.8\linewidth]{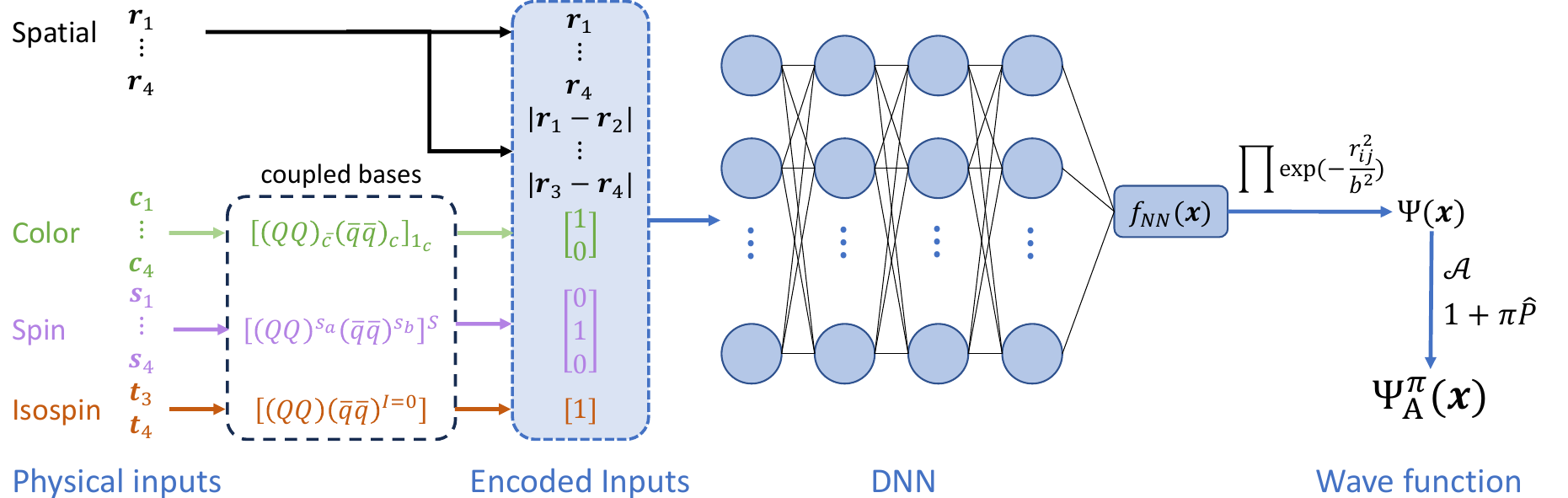}
		\caption{Architecture of the DeepQuark wave function, taking isoscalar doubly heavy tetraquark as an example. The physical inputs in spatial, color, spin and isospin degrees of freedom are transformed as the encoded inputs before fed into the DNN. Boundary condition, fermionic antisymmetrization ($\mathcal{A}$) and parity projection $(1+\pi\hat{P})$ are imposed on the scalar output of the DNN $f_{NN}(\boldsymbol{x})$ to obtain the multiquark wave function $\Psi_{\text{A}}^\pi(\boldsymbol{x})$. }
		\label{fig:frame}
	\end{figure*}

	\emph{Neural-network wave function}---The architecture of the DeepQuark wave function is illustrated in Fig.~\ref{fig:frame}. The core of the framework is to construct a multiquark wave function with the desired symmetry, namely, a color-singlet state with definite total spin $S$, isospin $I$ and parity $\pi$ that obeys Fermi-Dirac statistics. To that end, we work in the coupled bases. Taking isoscalar vector doubly heavy tetraquark as an example, the color-spin-isospin degrees of freedom can be coupled into the following independent bases:
	\begin{equation}
		\begin{aligned}
			&\chi_{\bar 3_c\otimes 3_c}\phi^{s_a,s_b}\xi^{I=0}=\left[\left(QQ\right)_{\bar 3_c}^{s_a}\left(\bar q\bar q\right)_{3_c}^{s_b,I=0}\right]_{1_c}^{S=1},\\
			&\chi_{6_c\otimes \bar6_c}\phi^{s_a,s_b}\xi^{I=0}=\left[\left(QQ\right)_{6_c}^{s_a}\left(\bar q\bar q\right)_{\bar6_c}^{s_b,I=0}\right]_{1_c}^{S=1},\\
		\end{aligned}
	\end{equation}
	where $s_a$ and $s_b$ take all possible combinations. $\chi$, $\phi$, and $\xi$ represent the color, spin, and isospin coupled bases, which are further mapped into vectors $\alpha_c$, $\alpha_s$ and $\alpha_t$, respectively. Here, $\alpha$ is a standard basis vector in $\mathbb{R}^n$, with $n$ being the number of independent bases. The mapping into standard basis vectors rather than integers from $1$ to $n$ ensures that no prejudiced correlation between different bases is introduced. By taking the vectors $\alpha$ as inputs of the DNN, the symmetry information is encoded into the DeepQuark wave function. 
	
	For the spatial degree of freedom, we include the (anti)quark coordinates in the center of mass frame $\boldsymbol{r}_i$ and the distances between two (anti)quarks $\left|\boldsymbol{r}_i-\boldsymbol{r}_j\right|$ as input features. The inclusion of the distances is redundant in principle but can improve the performance of neural-network wave function~\cite{FermiNet}. The magnitude of interparticle distance is a nonsmooth function at zero, which could better describe the wave function cusps~\cite{FermiNet} arising from short-range color Coulomb interaction  in Eq.~\eqref{eq:AL1}. It is also an important variable with direct physical meaning, which can encapsulate interparticle correlations effectively. As the wave function is a function of three-dimensional coordinates $\boldsymbol{r}_i$, it includes contributions from all  orbital angular momentum $L$. The wave function should naturally converge to the S-wave ground state in the optimization process. \clabel[B1-2]{It is also worth noting that the DeepQuark wave function ansatz does not encode any \textit{a priori} bias toward either molecular or compact states, making it structurally neutral with respect to multiquark configurations.}
	
	The input features $\boldsymbol{x}=\left(\boldsymbol{r}_i,\left|\boldsymbol{r}_i-\boldsymbol{r}_j\right|,\alpha_c,\alpha_s,\alpha_t\right)$ are fed into a DNN with four fully connected hidden layers. Each layer takes the outputs of the previous layer as inputs and carries out the mapping $\boldsymbol{x}_{\text{out}}=\sigma(\boldsymbol{W}\boldsymbol{x}_{\text{in}}+\boldsymbol{c})$,
	where $\boldsymbol{W}$, $\boldsymbol{c}$ are variational parameters, and $\sigma=\tanh$ is the activation function. \clabel[B2]{Since the parameters are initialized randomly, the initial trial wave function may be broadly distributed in the configuration space. To generate an initial state closer to the ground-state solution for faster convergence, we multiply the scalar output of the network $f_{NN}$ by a boundary factor $\prod_{i<j}\exp(-r_{ij}^2/b^2)$, which constrains the system to a localized region. We take $b=2-4$ fm, which is on the order of the typical range of color confinement, $\Lambda_{\text{QCD}}^{-1}\sim1$ fm. We have verified that the ground-state results are independent of the boundary parameter $b$.} Finally, the fermionic antisymmetry and parity projection is enforced,
	\begin{equation}
		\Psi_{\text{A}}^\pi(\boldsymbol{x})=(1+\pi\hat{P})\mathcal{A}\left[f_{NN}(\boldsymbol{x})\prod_{i<j}\exp\left(-\frac{r_{ij}^2}{b^2}\right)\right],
	\end{equation}
	where $\hat{P}$ and $\mathcal{A}$ are the spatial inversion operator and antisymmetric operator of identical particles, respectively. Enforcing antisymmetry by explicitly summing over all possible permutations leads to a factorial complexity. However, given the current experimental and theoretical progress, we will only focus on multiquark systems with at most three to four identical particles, and such a complexity is manageable.
	
	The DNN is trained in an unsupervised way using the variational principle. A good enough wave function ansatz converges to the ground state in the optimization process. More details can be found in the Supplemental Material~\cite{Suppl}.
	
	\emph{Results and discussion}---As a warm-up exercise, we test DeepQuark on few-electron systems, including $e^+e^-$ ($\mathrm{Ps}$), $e^+e^-e^-$ ($\mathrm{Ps}^-$) and $e^+e^+e^-e^-$ ($\mathrm{Ps}_2$). They can be regarded as the quantum electrodynamics counterparts of multiquark systems~\cite{Ma:2025rvj}. The DeepQuark results can reach a high accuracy with less than $1\text{\textperthousand}$ relative difference compared to the benchmark energies~(see Supplemental Material~\cite{Suppl}). 
    
    \begin{figure*}[tbp]
        \centering
        \includegraphics[width=\linewidth]{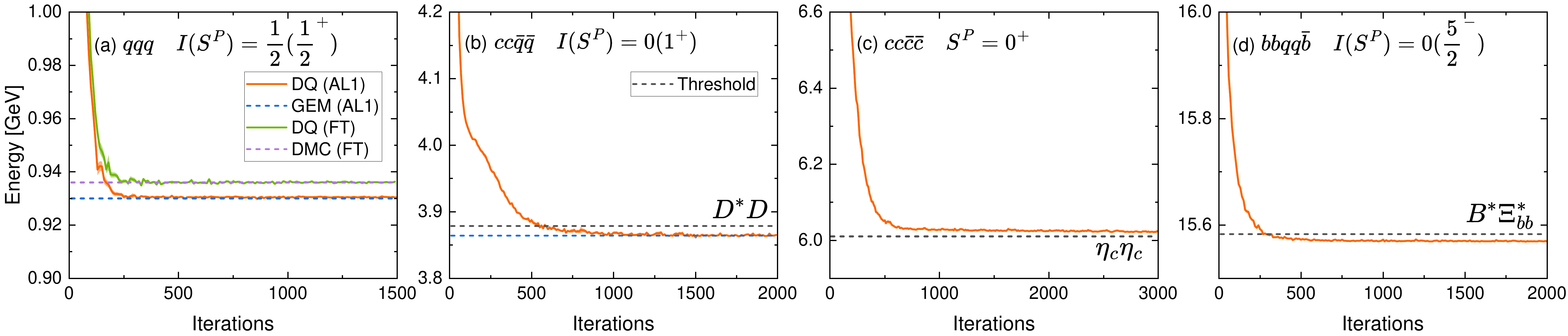}
        \caption{The energy estimate as a function of iteration steps for the (a) nucleon in the AL1 potential and flux-tube confinement interaction (FT), (b) isoscalar vector doubly charmed tetraquark, (c) scalar fully charmed tetraquark, and (d) isoscalar triply bottomed tetraquark systems in the optimization progress of DeepQuark (DQ). The Monte Carlo standard errors of the energies are shown by the shaded area, which are very tiny. The lowest two-body dissociation thresholds are represented by the black dashed lines. The ground-state energies given by the GEM~\cite{Ma:2022vqf,Meng:2023jqk} and DMC~\cite{Ma:2022vqf} are respectively displayed by the blue and purple dashed lines for comparison. }
        \label{fig:multi}
    \end{figure*}
    
    The optimization performance of DeepQuark in hadron systems is shown in Fig.~\ref{fig:multi}. We take four systems as examples, including the nucleon, doubly charmed and fully charmed tetraquarks, and triply bottomed pentaquark. It can be seen that DeepQuark achieves good convergence within a few thousand iterations for all these systems. \clabel[B3]{The statistical uncertainties (i.e., the Monte Carlo standard errors shown in Fig.~\ref{fig:multi}) of the DeepQuark energy results are below $0.1$ MeV, which are significantly smaller than the model uncertainty and can therefore be neglected.} In the nucleon system, DeepQuark quickly converges to the benchmark results from GEM and DMC~\cite{Ma:2022vqf} for both two-body AL1 interaction in Eq.~\eqref{eq:AL1} and flux-tube confinement interaction.  Calculating flux-tube interactions is computationally intractable in basis expansion methods, whereas DeepQuark handles them efficiently through Monte Carlo evaluation. This highlights DeepQuark's ability in managing complicate potentials of various forms.

	For tetraquark states, we investigate two systems that are of most concern, the isoscalar doubly heavy tetraquark $QQ\bar q\bar q\,(T_{QQ})$ and fully heavy tetraquark $QQ\bar Q\bar Q\,(T_{4Q})$. The existence of a deeply bound $T_{QQ}$ state for large enough heavy quark mass has long been anticipated~\cite{Zouzou1986,Manohar:1992nd} and was also predicted by lattice QCD study~\cite{Francis:2016hui,Junnarkar:2018twb}. Moreover, great experimental progress has been made in the discovery of $T_{cc}$~\cite{LHCb:2021vvq,LHCb:2021auc}. The $T_{4Q}$ system serves as a clear platform to investigate the short-range gluon exchange and confinement interaction, as it is less affected by the chiral dynamics. Bound states in such a system are inconclusive but a family of $T_{4c}$ resonances have been discovered experimentally~\cite{LHCb:2020bwg,CMS:2023owd,ATLAS:2023bft}. Recently, the CMS collaboration determined three $T_{4c}$ resonances with high significances and analyzed their quantum numbers to be $J^{PC}=2^{++}$~\cite{CMS:2025fpt}.  In Table~\ref{tab:tetra}, we present the ground-state properties of these tetraquark states and compare them with the results from GEM, which was shown to be a superior numerical approach to solving tetraquark bound states~\cite{Meng:2023jqk}.  \clabel[B4]{For $T_{QQ}$ systems, the DeepQuark ground-state energies are consistent with the GEM results in general, and lower the energies by $\simeq1$ MeV (see Table~\ref{tab:tetra}). This shows that DeepQuark has a stronger expressive power for the multiquark wave function than GEM, as indicated by the variational principle.} The ground state of $T_{bb}$ is dominated by the $\chi_{\bar 3_c\otimes 3_c}$ color configuration, whereas $T_{cc}$ exhibits sizable contributions from both $\chi_{\bar 3_c\otimes 3_c}$ and $\chi_{6_c\otimes \bar 6_c}$ configurations. Starting from a randomly initialized trial wave function, DeepQuark naturally converges to these two ground states with distinct mixing effects, demonstrating its strong capability in handling coupled-channel problems. On the other hand, no bound state solution is found in $T_{4Q}$ systems. The ground-state energies lie above the lowest meson-meson thresholds, and the color proportions $\chi_{\bar 3_c\otimes 3_c}:\chi_{6_c\otimes \bar6_c}\simeq1:2$ are in accordance with the meson-meson scattering states~\cite{Wu:2024euj}. However, since the DeepQuark wave function is constrained in a localized space, it cannot converge to an ideal scattering state with zero relative momentum. Therefore, the ground-state energies are $\sim 10$ MeV above the thresholds and the color proportions slightly deviates from $\chi_{\bar 3_c\otimes 3_c}:\chi_{6_c\otimes \bar6_c}\simeq1:2$.

    \begin{table*}[tbp]
		\centering
		\caption{Ground-state energies, color proportions and rms radii (in fm) of the isoscalar doubly heavy ($QQ\bar q\bar q$) and fully heavy $(QQ\bar Q\bar Q)$ tetraquark systems. The binding energies $\Delta E$ (in MeV) are with respect to the lowest dihadron thresholds listed in the third column. ``NB" indicates no bound state solution. The statistical errors of the energies from DeepQuark (DQ) are less than 0.1 MeV. The ground-state energies from GEM~\cite{Meng:2023jqk} are listed for comparison.}
		\label{tab:tetra}
		\begin{tabular*}{\hsize}{@{}@{\extracolsep{\fill}}lccccccccc@{}}
			\hline\hline
			&&&\multicolumn{6}{c}{DQ}&GEM~\cite{Meng:2023jqk}\\
			\hline
			&$S^P$&Thresholds&$\Delta E$&$ \chi_{\bar3_c\otimes 3_c} $&$ \chi_{6_c\otimes\bar 6_c} $&$r_{Q\bar q}$&$r_{QQ}$&$r_{\bar q\bar q}$&$\Delta E$\\
			\hline
			$cc\bar q\bar q$&$1^+$&$D^*D$&-15&55\%&45\%&1.06&1.24&1.41&-14\\
			$bb\bar q\bar q$&$1^+$&$\bar B^*\bar B$&-153&97\%&3\%&0.69&0.33&0.78&-152\\
			$cc\bar c\bar c$&$0^+,1^+,2^+$&$\eta_c\eta_c,\eta_cJ/\psi,J/\psi J/\psi$&NB&$\sim$34\%&$\sim$66\%&&&&NB\\
			$bb\bar b\bar b$&$0^+,1^+,2^+$&$\eta_b\eta_b,\eta_b\Upsilon,\Upsilon \Upsilon$&NB&$\sim$34\%&$\sim$66\%&&&&NB\\
			\hline\hline
		\end{tabular*}
	\end{table*}
    
	The DeepQuark wave function can consistently describe multiquark states with various configurations. \clabel[B1-3]{It is capabale of capturing the dynamical effects in different systems, converging to a compact $T_{bb}$ state and a molecular $T_{cc}$ state.} The root-mean-square (rms) radii of $T_{bb}$ and $T_{cc}$ are shown in Table~\ref{tab:tetra}. $T_{bb}$ is a distinct compact tetraquark, where two bottom quarks form a compact diquark cluster of $0.33$ fm and the light quarks orbit around the heavy diquark. \clabel[A2-1]{It is worth noting that a deeply bound $T_{bb}$ state with a binding energy of around $150$ MeV is in agreement with lattice QCD results, see~\cite{Francis:2016hui,Junnarkar:2018twb}}. In contrast to the compact $T_{bb}$, the size of $T_{cc}$ is significantly larger, with $r_{cc},r_{\bar q \bar q}>r_{c\bar q}$, suggesting a molecular structure of two charmed mesons~\cite{Li:2012ss}.  A loosely bound molecular $T_{cc}$ is in agreement with the experimental results~\cite{LHCb:2021vvq,LHCb:2021auc}. The separation between two charmed mesons becomes larger when the binding energy is tuned to be as small as the experimental one~\cite{Wu:2024zbx}.

	Inspired by the existence of $\bar Q\bar Qqq$ bound states,  we further employ DeepQuark to investigate the isoscalar triply heavy pentaquark $(QQqq\bar Q)$, which is the partner of the doubly heavy tetraquark if we consider the diquark-antiquark symmetry~\cite{Guo:2013xga,Wang:2024yjp}, namely replacing a heavy antiquark $\bar Q$ by a heavy diquark $QQ$. As the two heavy quarks may form a compact object with the same color representation $\bar 3_c$ as the antiquark, it is expected that the interactions in these two systems are similar, and $\bar Q\bar Qqq$ bound states may imply the existence of $QQqq\bar Q$ bound states.   DeepQuark has great advantages in extending to such pentaquark systems, since the computational complexity is almost the same as solving the tetraquark systems.  A complete set of color basis for the pentaquark is given by
	\begin{equation}
		\begin{aligned}
			\chi_{\bar3_c\otimes\bar 3_c}=\left\{\left[(QQ)_{\bar 3_c}(qq)_{\bar 3_c}\right]_{3_c}\bar Q\right\}_{1_c},\\
			\chi_{\bar3_c\otimes6_c}=\left\{\left[(QQ)_{\bar 3_c}(qq)_{6_c}\right]_{3_c}\bar Q\right\}_{1_c},\\
			\chi_{6_c\otimes\bar3_c}=\left\{\left[(QQ)_{6_c}(qq)_{\bar 3_c}\right]_{3_c}\bar Q\right\}_{1_c}.\\
		\end{aligned}
	\end{equation}

    \begin{table*}[!htp]
		\centering
		\caption{Ground-state energies, color  proportions and rms radii (in fm) of the isoscalar triply heavy pentaquark $(QQqq\bar Q)$ systems. The remaining notations follow the same conventions as in Table~\ref{tab:tetra}.}
		\label{tab:penta}
		\begin{tabular*}{\hsize}{@{}@{\extracolsep{\fill}}lccccccccccc@{}}
			\hline\hline
			&$S^P$&Thresholds&$\Delta E$&$ \chi_{\bar3_c\otimes\bar 3_c} $&$ \chi_{\bar3_c\otimes 6_c} $&$ \chi_{6_c\otimes \bar3_c} $&$r_{QQ}$&$r_{Qq}$&$r_{qq}$&$r_{Q\bar Q}$& $r_{q\bar Q}$\\
			\hline
			$ccqq\bar c$&$\frac{1}{2}^-,\frac{3}{2}^-$&$\eta_c\Lambda_c,J/\psi\Lambda_c$&NB&$\sim$35\%&0\%&$\sim$65\%&&&&&\\
			&$\frac{5}{2}^-$&$\bar D^*\Xi_{cc}^*$&-3&27\%&73\%&0\%&0.50&1.39&1.90&1.73&1.38\\
			$bbqq\bar b$&$\frac{1}{2}^-,\frac{3}{2}^-$&$\eta_b\Lambda_b,\Upsilon\Lambda_b$&NB&$\sim$35\%&0\%&$\sim$65\%&&&&&\\
			&$\frac{5}{2}^-$&$B^*\Xi_{bb}^*$&-14&19\%&80\%&1\%&0.30&0.89&1.22&0.88&0.88	\\
			\hline\hline
		\end{tabular*}
        
	\end{table*}
	We show the ground-state properties in Table~\ref{tab:penta}. In contrast to the expectations from diquark-antiquark symmetry, we find no bound state in the $S=\frac{1}{2},\frac{3}{2}$ $QQqq\bar Q$ systems. The reason is that the existence of $\bar Q$ breaks the diquark-antiquark symmetry by taking one of the heavy quarks to form a quarkonium $(Q\bar Q)_{1_c}$, which is more stable and has lower energy than the configuration with $(QQ)_{\bar 3_c}$. The ground states in these systems are scattering states of a heavy quarkonium and a singly heavy baryon. However, in $S=\frac{5}{2}$ system, an S-wave $S=\frac{3}{2}$ singly heavy isoscalar baryon is not allowed by the antisymmetry of light quarks. As a result, the lowest threshold is $\bar D^*\Xi_{cc}^*$ for $ccqq\bar c$ and $B^*\Xi_{bb}^*$ for $bbqq\bar b$. We find bound state solutions $P_{cc\bar c}(5715)$ and $P_{bb\bar b}(15569)$, whose binding energies are $3$ MeV and $14$ MeV, respectively. They may decay to meson-baryon channels with lower total spin if spin-orbit coupling is considered. For example, $P_{cc\bar c}(5715)$ may be searched for in the $J/\psi\Lambda_c$ channel, but its width is expected to be suppressed by the D-wave decay. From the rms radii of $P_{cc\bar c}(5715)$, the distance $r_{cc}=0.5$ fm is consistent with the compact size of $\Xi_{cc}^*$, whereas $r_{c\bar c}=1.73$ fm is substantially larger. The spatial separation suggests a molecular configuration composed of loosely bound $\bar D^*$ and $\Xi_{cc}^*$, which is analogous to the molecular $T_{cc}$. $P_{bb\bar b}(15569)$ exhibits a similar configuration, but its size is compacted by the heavy quark mass.

    \emph{Conclusions and outlooks}---We introduce a novel and high-efficiency DNN-based VMC approach, DeepQuark, to solving the multiquark bound states. We exploit the expressive power of DNN to construct an unbiased and symmetry-informed wave function ansatz, which is capable of handling complex interactions and describing various multiquark configurations.  DeepQuark is competitive with state-of-the-art approaches including DMC and GEM in baryon and tetraquark systems, reaching high accuracies on ground-state energies with low computational costs. In the nucleon, DeepQuark easily adapts both two-body and flux-tube interactions. In tetraquark systems, it consistently describes meson molecule $T_{cc}$ and compact tetraquark $T_{bb}$. Moreover, we employ DeepQuark to investigate triply heavy pentaquark ($QQqq\bar Q$) systems. We obtain weakly bound $\bar D^*\Xi_{cc}^*$ molecule $P_{cc\bar c}(5715)$ with $S=\frac{5}{2}$ and its bottom partner $P_{bb\bar b}(15569)$. They can be viewed as the analogs of the molecular $T_{cc}$. \clabel[A6]{We recommend experimental search for $P_{cc\bar c}(5715)$ in the D-wave $J/\psi \Lambda_c$ channel, which could be explored in future LHCb experiments.} 
    
    DeepQuark holds great promise for extension to diverse pentaquark and even hexaquark systems, overcoming the computational barriers that limit conventional methods. Such investigations can provide forward-looking predictions for future experiments. \clabel[A2-2]{Furthermore, DeepQuark is capable of incorporating a variety of potential models with different dynamical ingredients, including complex interactions beyond two-body forces. It therefore provides a powerful framework for investigating lattice-inspired potentials and exploring the confining mechanisms of tetraquark states in future studies, potentially offering valuable insights into the multiquark internal structures and nonperturbative QCD.} The techniques employed in DeepQuark could enrich the toolkit of deep learning approaches integrated with physics, particularly shedding new light on quantum many-body phenomena. 
    
    \emph{Acknowledgments}---We thank Yao Ma, Yan-Ke Chen, and Liang-Zhen Wen for the helpful discussions. L.M. also thanks Yilong Yang and Pengwei Zhao for discussions on DNN and VMC.  This project was supported by the National
    Natural Science Foundation of China (Grant No. 12475137 and No.125B2105), ERC NuclearTheory (Grant No. 885150) and  Start-up Funds of Southeast University (Grant No. 4007022506). The codes are developed on top of JAX~\cite{jax2018github}, Flax~\cite{flax2020github}, and the NetKet package~\cite{netket3:2022,netket2:2019}. The computational resources are supported by High-performance Computing Platform of Peking University.

  \emph{Data Availability}---The data supporting this study’s findings are available
 within the article. \clabel[code]{The codes that can reproduce the reported results are available on GitHub~\cite{Wu_DeepQuark_2026}}.
 
\bibliography{DQRef}

\begin{thebibliography}{87}%
\makeatletter
\providecommand \@ifxundefined [1]{%
 \@ifx{#1\undefined}
}%
\providecommand \@ifnum [1]{%
 \ifnum #1\expandafter \@firstoftwo
 \else \expandafter \@secondoftwo
 \fi
}%
\providecommand \@ifx [1]{%
 \ifx #1\expandafter \@firstoftwo
 \else \expandafter \@secondoftwo
 \fi
}%
\providecommand \natexlab [1]{#1}%
\providecommand \enquote  [1]{``#1''}%
\providecommand \bibnamefont  [1]{#1}%
\providecommand \bibfnamefont [1]{#1}%
\providecommand \citenamefont [1]{#1}%
\providecommand \href@noop [0]{\@secondoftwo}%
\providecommand \href [0]{\begingroup \@sanitize@url \@href}%
\providecommand \@href[1]{\@@startlink{#1}\@@href}%
\providecommand \@@href[1]{\endgroup#1\@@endlink}%
\providecommand \@sanitize@url [0]{\catcode `\\12\catcode `\$12\catcode
  `\&12\catcode `\#12\catcode `\^12\catcode `\_12\catcode `\%12\relax}%
\providecommand \@@startlink[1]{}%
\providecommand \@@endlink[0]{}%
\providecommand \url  [0]{\begingroup\@sanitize@url \@url }%
\providecommand \@url [1]{\endgroup\@href {#1}{\urlprefix }}%
\providecommand \urlprefix  [0]{URL }%
\providecommand \Eprint [0]{\href }%
\providecommand \doibase [0]{https://doi.org/}%
\providecommand \selectlanguage [0]{\@gobble}%
\providecommand \bibinfo  [0]{\@secondoftwo}%
\providecommand \bibfield  [0]{\@secondoftwo}%
\providecommand \translation [1]{[#1]}%
\providecommand \BibitemOpen [0]{}%
\providecommand \bibitemStop [0]{}%
\providecommand \bibitemNoStop [0]{.\EOS\space}%
\providecommand \EOS [0]{\spacefactor3000\relax}%
\providecommand \BibitemShut  [1]{\csname bibitem#1\endcsname}%
\let\auto@bib@innerbib\@empty
\bibitem [{\citenamefont {Gell-Mann}(1964)}]{Gell-Mann:1964ewy}%
  \BibitemOpen
  \bibfield  {author} {\bibinfo {author} {\bibfnamefont {M.}~\bibnamefont
  {Gell-Mann}},\ }\bibfield  {title} {\bibinfo {title} {{A Schematic Model of
  Baryons and Mesons}},\ }\href {https://doi.org/10.1016/S0031-9163(64)92001-3}
  {\bibfield  {journal} {\bibinfo  {journal} {Phys. Lett.}\ }\textbf {\bibinfo
  {volume} {8}},\ \bibinfo {pages} {214} (\bibinfo {year} {1964})}\BibitemShut
  {NoStop}%
\bibitem [{\citenamefont {Zweig}(1964)}]{Zweig:1964ruk}%
  \BibitemOpen
  \bibfield  {author} {\bibinfo {author} {\bibfnamefont {G.}~\bibnamefont
  {Zweig}},\ }\bibfield  {title} {\bibinfo {title} {{An SU(3) model for strong
  interaction symmetry and its breaking. Version 1}}\ }\href
  {https://doi.org/10.17181/CERN-TH-401} {10.17181/CERN-TH-401} (\bibinfo
  {year} {1964})\BibitemShut {NoStop}%
\bibitem [{\citenamefont {Choi}\ \emph {et~al.}(2003)\citenamefont {Choi} \emph
  {et~al.}}]{Belle:2003nnu}%
  \BibitemOpen
  \bibfield  {author} {\bibinfo {author} {\bibfnamefont {S.~K.}\ \bibnamefont
  {Choi}} \emph {et~al.} (\bibinfo {collaboration} {Belle}),\ }\bibfield
  {title} {\bibinfo {title} {{Observation of a narrow charmonium-like state in
  exclusive $B^\pm \to K^\pm \pi^+ \pi^- J/\psi$ decays}},\ }\href
  {https://doi.org/10.1103/PhysRevLett.91.262001} {\bibfield  {journal}
  {\bibinfo  {journal} {Phys. Rev. Lett.}\ }\textbf {\bibinfo {volume} {91}},\
  \bibinfo {pages} {262001} (\bibinfo {year} {2003})},\ \Eprint
  {https://arxiv.org/abs/hep-ex/0309032} {arXiv:hep-ex/0309032} \BibitemShut
  {NoStop}%
\bibitem [{\citenamefont {Ablikim}\ \emph {et~al.}(2013)\citenamefont {Ablikim}
  \emph {et~al.}}]{BESIII:2013ris}%
  \BibitemOpen
  \bibfield  {author} {\bibinfo {author} {\bibfnamefont {M.}~\bibnamefont
  {Ablikim}} \emph {et~al.} (\bibinfo {collaboration} {BESIII}),\ }\bibfield
  {title} {\bibinfo {title} {{Observation of a Charged Charmoniumlike Structure
  in $e^+e^- \to \pi^+\pi^- J/\psi$ at $\sqrt{s}$ =4.26 GeV}},\ }\href
  {https://doi.org/10.1103/PhysRevLett.110.252001} {\bibfield  {journal}
  {\bibinfo  {journal} {Phys. Rev. Lett.}\ }\textbf {\bibinfo {volume} {110}},\
  \bibinfo {pages} {252001} (\bibinfo {year} {2013})},\ \Eprint
  {https://arxiv.org/abs/1303.5949} {arXiv:1303.5949 [hep-ex]} \BibitemShut
  {NoStop}%
\bibitem [{\citenamefont {Liu}\ \emph {et~al.}(2013)\citenamefont {Liu} \emph
  {et~al.}}]{Belle:2013yex}%
  \BibitemOpen
  \bibfield  {author} {\bibinfo {author} {\bibfnamefont {Z.~Q.}\ \bibnamefont
  {Liu}} \emph {et~al.} (\bibinfo {collaboration} {Belle}),\ }\bibfield
  {title} {\bibinfo {title} {{Study of $e^+e^- → \pi^+ \pi^- J/\psi$ and
  Observation of a Charged Charmoniumlike State at Belle}},\ }\href
  {https://doi.org/10.1103/PhysRevLett.110.252002} {\bibfield  {journal}
  {\bibinfo  {journal} {Phys. Rev. Lett.}\ }\textbf {\bibinfo {volume} {110}},\
  \bibinfo {pages} {252002} (\bibinfo {year} {2013})},\ \bibinfo {note}
  {[Erratum: Phys.Rev.Lett. 111, 019901 (2013)]},\ \Eprint
  {https://arxiv.org/abs/1304.0121} {arXiv:1304.0121 [hep-ex]} \BibitemShut
  {NoStop}%
\bibitem [{\citenamefont {Aaij}\ \emph
  {et~al.}(2022{\natexlab{a}})\citenamefont {Aaij} \emph
  {et~al.}}]{LHCb:2021vvq}%
  \BibitemOpen
  \bibfield  {author} {\bibinfo {author} {\bibfnamefont {R.}~\bibnamefont
  {Aaij}} \emph {et~al.} (\bibinfo {collaboration} {LHCb}),\ }\bibfield
  {title} {\bibinfo {title} {{Observation of an exotic narrow doubly charmed
  tetraquark}},\ }\href {https://doi.org/10.1038/s41567-022-01614-y} {\bibfield
   {journal} {\bibinfo  {journal} {Nature Phys.}\ }\textbf {\bibinfo {volume}
  {18}},\ \bibinfo {pages} {751} (\bibinfo {year} {2022}{\natexlab{a}})},\
  \Eprint {https://arxiv.org/abs/2109.01038} {arXiv:2109.01038 [hep-ex]}
  \BibitemShut {NoStop}%
\bibitem [{\citenamefont {Aaij}\ \emph
  {et~al.}(2022{\natexlab{b}})\citenamefont {Aaij} \emph
  {et~al.}}]{LHCb:2021auc}%
  \BibitemOpen
  \bibfield  {author} {\bibinfo {author} {\bibfnamefont {R.}~\bibnamefont
  {Aaij}} \emph {et~al.} (\bibinfo {collaboration} {LHCb}),\ }\bibfield
  {title} {\bibinfo {title} {{Study of the doubly charmed tetraquark
  $T_{cc}^{+}$}},\ }\href {https://doi.org/10.1038/s41467-022-30206-w}
  {\bibfield  {journal} {\bibinfo  {journal} {Nature Commun.}\ }\textbf
  {\bibinfo {volume} {13}},\ \bibinfo {pages} {3351} (\bibinfo {year}
  {2022}{\natexlab{b}})},\ \Eprint {https://arxiv.org/abs/2109.01056}
  {arXiv:2109.01056 [hep-ex]} \BibitemShut {NoStop}%
\bibitem [{\citenamefont {Aaij}\ \emph {et~al.}(2015)\citenamefont {Aaij} \emph
  {et~al.}}]{LHCb:2015yax}%
  \BibitemOpen
  \bibfield  {author} {\bibinfo {author} {\bibfnamefont {R.}~\bibnamefont
  {Aaij}} \emph {et~al.} (\bibinfo {collaboration} {LHCb}),\ }\bibfield
  {title} {\bibinfo {title} {{Observation of $J/\psi p$ Resonances Consistent
  with Pentaquark States in $\Lambda_b^0 \to J/\psi K^- p$ Decays}},\ }\href
  {https://doi.org/10.1103/PhysRevLett.115.072001} {\bibfield  {journal}
  {\bibinfo  {journal} {Phys. Rev. Lett.}\ }\textbf {\bibinfo {volume} {115}},\
  \bibinfo {pages} {072001} (\bibinfo {year} {2015})},\ \Eprint
  {https://arxiv.org/abs/1507.03414} {arXiv:1507.03414 [hep-ex]} \BibitemShut
  {NoStop}%
\bibitem [{\citenamefont {Aaij}\ \emph {et~al.}(2019)\citenamefont {Aaij} \emph
  {et~al.}}]{LHCb:2019kea}%
  \BibitemOpen
  \bibfield  {author} {\bibinfo {author} {\bibfnamefont {R.}~\bibnamefont
  {Aaij}} \emph {et~al.} (\bibinfo {collaboration} {LHCb}),\ }\bibfield
  {title} {\bibinfo {title} {{Observation of a narrow pentaquark state,
  $P_c(4312)^+$, and of two-peak structure of the $P_c(4450)^+$}},\ }\href
  {https://doi.org/10.1103/PhysRevLett.122.222001} {\bibfield  {journal}
  {\bibinfo  {journal} {Phys. Rev. Lett.}\ }\textbf {\bibinfo {volume} {122}},\
  \bibinfo {pages} {222001} (\bibinfo {year} {2019})},\ \Eprint
  {https://arxiv.org/abs/1904.03947} {arXiv:1904.03947 [hep-ex]} \BibitemShut
  {NoStop}%
\bibitem [{\citenamefont {Aaij}\ \emph {et~al.}(2020)\citenamefont {Aaij} \emph
  {et~al.}}]{LHCb:2020bwg}%
  \BibitemOpen
  \bibfield  {author} {\bibinfo {author} {\bibfnamefont {R.}~\bibnamefont
  {Aaij}} \emph {et~al.} (\bibinfo {collaboration} {LHCb}),\ }\bibfield
  {title} {\bibinfo {title} {{Observation of structure in the $J /\psi$ -pair
  mass spectrum}},\ }\href {https://doi.org/10.1016/j.scib.2020.08.032}
  {\bibfield  {journal} {\bibinfo  {journal} {Sci. Bull.}\ }\textbf {\bibinfo
  {volume} {65}},\ \bibinfo {pages} {1983} (\bibinfo {year} {2020})},\ \Eprint
  {https://arxiv.org/abs/2006.16957} {arXiv:2006.16957 [hep-ex]} \BibitemShut
  {NoStop}%
\bibitem [{\citenamefont {Chen}\ \emph {et~al.}(2016)\citenamefont {Chen},
  \citenamefont {Chen}, \citenamefont {Liu},\ and\ \citenamefont
  {Zhu}}]{Chen:2016qju}%
  \BibitemOpen
  \bibfield  {author} {\bibinfo {author} {\bibfnamefont {H.-X.}\ \bibnamefont
  {Chen}}, \bibinfo {author} {\bibfnamefont {W.}~\bibnamefont {Chen}}, \bibinfo
  {author} {\bibfnamefont {X.}~\bibnamefont {Liu}},\ and\ \bibinfo {author}
  {\bibfnamefont {S.-L.}\ \bibnamefont {Zhu}},\ }\bibfield  {title} {\bibinfo
  {title} {{The hidden-charm pentaquark and tetraquark states}},\ }\href
  {https://doi.org/10.1016/j.physrep.2016.05.004} {\bibfield  {journal}
  {\bibinfo  {journal} {Phys. Rept.}\ }\textbf {\bibinfo {volume} {639}},\
  \bibinfo {pages} {1} (\bibinfo {year} {2016})},\ \Eprint
  {https://arxiv.org/abs/1601.02092} {arXiv:1601.02092 [hep-ph]} \BibitemShut
  {NoStop}%
\bibitem [{\citenamefont {Hosaka}\ \emph {et~al.}(2016)\citenamefont {Hosaka},
  \citenamefont {Iijima}, \citenamefont {Miyabayashi}, \citenamefont {Sakai},\
  and\ \citenamefont {Yasui}}]{Hosaka:2016pey}%
  \BibitemOpen
  \bibfield  {author} {\bibinfo {author} {\bibfnamefont {A.}~\bibnamefont
  {Hosaka}}, \bibinfo {author} {\bibfnamefont {T.}~\bibnamefont {Iijima}},
  \bibinfo {author} {\bibfnamefont {K.}~\bibnamefont {Miyabayashi}}, \bibinfo
  {author} {\bibfnamefont {Y.}~\bibnamefont {Sakai}},\ and\ \bibinfo {author}
  {\bibfnamefont {S.}~\bibnamefont {Yasui}},\ }\bibfield  {title} {\bibinfo
  {title} {{Exotic hadrons with heavy flavors: X, Y, Z, and related states}},\
  }\href {https://doi.org/10.1093/ptep/ptw045} {\bibfield  {journal} {\bibinfo
  {journal} {PTEP}\ }\textbf {\bibinfo {volume} {2016}},\ \bibinfo {pages}
  {062C01} (\bibinfo {year} {2016})},\ \Eprint
  {https://arxiv.org/abs/1603.09229} {arXiv:1603.09229 [hep-ph]} \BibitemShut
  {NoStop}%
\bibitem [{\citenamefont {Lebed}\ \emph {et~al.}(2017)\citenamefont {Lebed},
  \citenamefont {Mitchell},\ and\ \citenamefont {Swanson}}]{Lebed:2016hpi}%
  \BibitemOpen
  \bibfield  {author} {\bibinfo {author} {\bibfnamefont {R.~F.}\ \bibnamefont
  {Lebed}}, \bibinfo {author} {\bibfnamefont {R.~E.}\ \bibnamefont
  {Mitchell}},\ and\ \bibinfo {author} {\bibfnamefont {E.~S.}\ \bibnamefont
  {Swanson}},\ }\bibfield  {title} {\bibinfo {title} {{Heavy-Quark QCD
  Exotica}},\ }\href {https://doi.org/10.1016/j.ppnp.2016.11.003} {\bibfield
  {journal} {\bibinfo  {journal} {Prog. Part. Nucl. Phys.}\ }\textbf {\bibinfo
  {volume} {93}},\ \bibinfo {pages} {143} (\bibinfo {year} {2017})},\ \Eprint
  {https://arxiv.org/abs/1610.04528} {arXiv:1610.04528 [hep-ph]} \BibitemShut
  {NoStop}%
\bibitem [{\citenamefont {Guo}\ \emph {et~al.}(2018)\citenamefont {Guo},
  \citenamefont {Hanhart}, \citenamefont {Mei\ss{}ner}, \citenamefont {Wang},
  \citenamefont {Zhao},\ and\ \citenamefont {Zou}}]{Guo:2017jvc}%
  \BibitemOpen
  \bibfield  {author} {\bibinfo {author} {\bibfnamefont {F.-K.}\ \bibnamefont
  {Guo}}, \bibinfo {author} {\bibfnamefont {C.}~\bibnamefont {Hanhart}},
  \bibinfo {author} {\bibfnamefont {U.-G.}\ \bibnamefont {Mei\ss{}ner}},
  \bibinfo {author} {\bibfnamefont {Q.}~\bibnamefont {Wang}}, \bibinfo {author}
  {\bibfnamefont {Q.}~\bibnamefont {Zhao}},\ and\ \bibinfo {author}
  {\bibfnamefont {B.-S.}\ \bibnamefont {Zou}},\ }\bibfield  {title} {\bibinfo
  {title} {{Hadronic molecules}},\ }\href
  {https://doi.org/10.1103/RevModPhys.90.015004} {\bibfield  {journal}
  {\bibinfo  {journal} {Rev. Mod. Phys.}\ }\textbf {\bibinfo {volume} {90}},\
  \bibinfo {pages} {015004} (\bibinfo {year} {2018})},\ \bibinfo {note}
  {[Erratum: Rev.Mod.Phys. 94, 029901 (2022)]},\ \Eprint
  {https://arxiv.org/abs/1705.00141} {arXiv:1705.00141 [hep-ph]} \BibitemShut
  {NoStop}%
\bibitem [{\citenamefont {Liu}\ \emph {et~al.}(2019)\citenamefont {Liu},
  \citenamefont {Chen}, \citenamefont {Chen}, \citenamefont {Liu},\ and\
  \citenamefont {Zhu}}]{Liu:2019zoy}%
  \BibitemOpen
  \bibfield  {author} {\bibinfo {author} {\bibfnamefont {Y.-R.}\ \bibnamefont
  {Liu}}, \bibinfo {author} {\bibfnamefont {H.-X.}\ \bibnamefont {Chen}},
  \bibinfo {author} {\bibfnamefont {W.}~\bibnamefont {Chen}}, \bibinfo {author}
  {\bibfnamefont {X.}~\bibnamefont {Liu}},\ and\ \bibinfo {author}
  {\bibfnamefont {S.-L.}\ \bibnamefont {Zhu}},\ }\bibfield  {title} {\bibinfo
  {title} {{Pentaquark and Tetraquark states}},\ }\href
  {https://doi.org/10.1016/j.ppnp.2019.04.003} {\bibfield  {journal} {\bibinfo
  {journal} {Prog. Part. Nucl. Phys.}\ }\textbf {\bibinfo {volume} {107}},\
  \bibinfo {pages} {237} (\bibinfo {year} {2019})},\ \Eprint
  {https://arxiv.org/abs/1903.11976} {arXiv:1903.11976 [hep-ph]} \BibitemShut
  {NoStop}%
\bibitem [{\citenamefont {Brambilla}\ \emph {et~al.}(2020)\citenamefont
  {Brambilla}, \citenamefont {Eidelman}, \citenamefont {Hanhart}, \citenamefont
  {Nefediev}, \citenamefont {Shen}, \citenamefont {Thomas}, \citenamefont
  {Vairo},\ and\ \citenamefont {Yuan}}]{Brambilla:2019esw}%
  \BibitemOpen
  \bibfield  {author} {\bibinfo {author} {\bibfnamefont {N.}~\bibnamefont
  {Brambilla}}, \bibinfo {author} {\bibfnamefont {S.}~\bibnamefont {Eidelman}},
  \bibinfo {author} {\bibfnamefont {C.}~\bibnamefont {Hanhart}}, \bibinfo
  {author} {\bibfnamefont {A.}~\bibnamefont {Nefediev}}, \bibinfo {author}
  {\bibfnamefont {C.-P.}\ \bibnamefont {Shen}}, \bibinfo {author}
  {\bibfnamefont {C.~E.}\ \bibnamefont {Thomas}}, \bibinfo {author}
  {\bibfnamefont {A.}~\bibnamefont {Vairo}},\ and\ \bibinfo {author}
  {\bibfnamefont {C.-Z.}\ \bibnamefont {Yuan}},\ }\bibfield  {title} {\bibinfo
  {title} {{The $XYZ$ states: experimental and theoretical status and
  perspectives}},\ }\href {https://doi.org/10.1016/j.physrep.2020.05.001}
  {\bibfield  {journal} {\bibinfo  {journal} {Phys. Rept.}\ }\textbf {\bibinfo
  {volume} {873}},\ \bibinfo {pages} {1} (\bibinfo {year} {2020})},\ \Eprint
  {https://arxiv.org/abs/1907.07583} {arXiv:1907.07583 [hep-ex]} \BibitemShut
  {NoStop}%
\bibitem [{\citenamefont {Chen}\ \emph {et~al.}(2023)\citenamefont {Chen},
  \citenamefont {Chen}, \citenamefont {Liu}, \citenamefont {Liu},\ and\
  \citenamefont {Zhu}}]{Chen:2022asf}%
  \BibitemOpen
  \bibfield  {author} {\bibinfo {author} {\bibfnamefont {H.-X.}\ \bibnamefont
  {Chen}}, \bibinfo {author} {\bibfnamefont {W.}~\bibnamefont {Chen}}, \bibinfo
  {author} {\bibfnamefont {X.}~\bibnamefont {Liu}}, \bibinfo {author}
  {\bibfnamefont {Y.-R.}\ \bibnamefont {Liu}},\ and\ \bibinfo {author}
  {\bibfnamefont {S.-L.}\ \bibnamefont {Zhu}},\ }\bibfield  {title} {\bibinfo
  {title} {{An updated review of the new hadron states}},\ }\href
  {https://doi.org/10.1088/1361-6633/aca3b6} {\bibfield  {journal} {\bibinfo
  {journal} {Rept. Prog. Phys.}\ }\textbf {\bibinfo {volume} {86}},\ \bibinfo
  {pages} {026201} (\bibinfo {year} {2023})},\ \Eprint
  {https://arxiv.org/abs/2204.02649} {arXiv:2204.02649 [hep-ph]} \BibitemShut
  {NoStop}%
\bibitem [{\citenamefont {Meng}\ \emph
  {et~al.}(2023{\natexlab{a}})\citenamefont {Meng}, \citenamefont {Wang},
  \citenamefont {Wang},\ and\ \citenamefont {Zhu}}]{Meng:2022ozq}%
  \BibitemOpen
  \bibfield  {author} {\bibinfo {author} {\bibfnamefont {L.}~\bibnamefont
  {Meng}}, \bibinfo {author} {\bibfnamefont {B.}~\bibnamefont {Wang}}, \bibinfo
  {author} {\bibfnamefont {G.-J.}\ \bibnamefont {Wang}},\ and\ \bibinfo
  {author} {\bibfnamefont {S.-L.}\ \bibnamefont {Zhu}},\ }\bibfield  {title}
  {\bibinfo {title} {{Chiral perturbation theory for heavy hadrons and chiral
  effective field theory for heavy hadronic molecules}},\ }\href
  {https://doi.org/10.1016/j.physrep.2023.04.003} {\bibfield  {journal}
  {\bibinfo  {journal} {Phys. Rept.}\ }\textbf {\bibinfo {volume} {1019}},\
  \bibinfo {pages} {1} (\bibinfo {year} {2023}{\natexlab{a}})},\ \Eprint
  {https://arxiv.org/abs/2204.08716} {arXiv:2204.08716 [hep-ph]} \BibitemShut
  {NoStop}%
\bibitem [{\citenamefont {Wang}(2026)}]{Wang:2025sic}%
  \BibitemOpen
  \bibfield  {author} {\bibinfo {author} {\bibfnamefont {Z.-G.}\ \bibnamefont
  {Wang}},\ }\bibfield  {title} {\bibinfo {title} {{Review of the QCD sum rules
  for exotic states}},\ }\href {https://doi.org/10.15302/frontphys.2026.016300}
  {\bibfield  {journal} {\bibinfo  {journal} {Front. Phys. (Beijing)}\ }\textbf
  {\bibinfo {volume} {21}},\ \bibinfo {pages} {016300} (\bibinfo {year}
  {2026})},\ \Eprint {https://arxiv.org/abs/2502.11351} {arXiv:2502.11351
  [hep-ph]} \BibitemShut {NoStop}%
\bibitem [{\citenamefont {Okiharu}\ \emph {et~al.}(2005)\citenamefont
  {Okiharu}, \citenamefont {Suganuma},\ and\ \citenamefont
  {Takahashi}}]{Okiharu:2004ve}%
  \BibitemOpen
  \bibfield  {author} {\bibinfo {author} {\bibfnamefont {F.}~\bibnamefont
  {Okiharu}}, \bibinfo {author} {\bibfnamefont {H.}~\bibnamefont {Suganuma}},\
  and\ \bibinfo {author} {\bibfnamefont {T.~T.}\ \bibnamefont {Takahashi}},\
  }\bibfield  {title} {\bibinfo {title} {{Detailed analysis of the tetraquark
  potential and flip-flop in SU(3) lattice QCD}},\ }\href
  {https://doi.org/10.1103/PhysRevD.72.014505} {\bibfield  {journal} {\bibinfo
  {journal} {Phys. Rev. D}\ }\textbf {\bibinfo {volume} {72}},\ \bibinfo
  {pages} {014505} (\bibinfo {year} {2005})},\ \Eprint
  {https://arxiv.org/abs/hep-lat/0412012} {arXiv:hep-lat/0412012} \BibitemShut
  {NoStop}%
\bibitem [{\citenamefont {Cardoso}\ \emph {et~al.}(2012)\citenamefont
  {Cardoso}, \citenamefont {Cardoso},\ and\ \citenamefont
  {Bicudo}}]{Cardoso:2012uka}%
  \BibitemOpen
  \bibfield  {author} {\bibinfo {author} {\bibfnamefont {M.}~\bibnamefont
  {Cardoso}}, \bibinfo {author} {\bibfnamefont {N.}~\bibnamefont {Cardoso}},\
  and\ \bibinfo {author} {\bibfnamefont {P.}~\bibnamefont {Bicudo}},\
  }\bibfield  {title} {\bibinfo {title} {{Variational study of the flux tube
  recombination in the two quarks and two quarks system in Lattice QCD}},\
  }\href {https://doi.org/10.1103/PhysRevD.86.014503} {\bibfield  {journal}
  {\bibinfo  {journal} {Phys. Rev. D}\ }\textbf {\bibinfo {volume} {86}},\
  \bibinfo {pages} {014503} (\bibinfo {year} {2012})},\ \Eprint
  {https://arxiv.org/abs/1204.5131} {arXiv:1204.5131 [hep-lat]} \BibitemShut
  {NoStop}%
\bibitem [{\citenamefont {Bicudo}\ and\ \citenamefont
  {Cardoso}(2016)}]{Bicudo:2015bra}%
  \BibitemOpen
  \bibfield  {author} {\bibinfo {author} {\bibfnamefont {P.}~\bibnamefont
  {Bicudo}}\ and\ \bibinfo {author} {\bibfnamefont {M.}~\bibnamefont
  {Cardoso}},\ }\bibfield  {title} {\bibinfo {title} {{Tetraquark bound states
  and resonances in the unitary and microscopic triple string flip-flop quark
  model, the light-light-antiheavy-antiheavy $q q \bar Q\bar Q$ case study}},\
  }\href {https://doi.org/10.1103/PhysRevD.94.094032} {\bibfield  {journal}
  {\bibinfo  {journal} {Phys. Rev. D}\ }\textbf {\bibinfo {volume} {94}},\
  \bibinfo {pages} {094032} (\bibinfo {year} {2016})},\ \Eprint
  {https://arxiv.org/abs/1509.04943} {arXiv:1509.04943 [hep-ph]} \BibitemShut
  {NoStop}%
\bibitem [{\citenamefont {Hiyama}\ \emph {et~al.}(2003)\citenamefont {Hiyama},
  \citenamefont {Kino},\ and\ \citenamefont {Kamimura}}]{Hiyama:2003cu}%
  \BibitemOpen
  \bibfield  {author} {\bibinfo {author} {\bibfnamefont {E.}~\bibnamefont
  {Hiyama}}, \bibinfo {author} {\bibfnamefont {Y.}~\bibnamefont {Kino}},\ and\
  \bibinfo {author} {\bibfnamefont {M.}~\bibnamefont {Kamimura}},\ }\bibfield
  {title} {\bibinfo {title} {{Gaussian expansion method for few-body
  systems}},\ }\href {https://doi.org/10.1016/S0146-6410(03)90015-9} {\bibfield
   {journal} {\bibinfo  {journal} {Prog. Part. Nucl. Phys.}\ }\textbf {\bibinfo
  {volume} {51}},\ \bibinfo {pages} {223} (\bibinfo {year} {2003})}\BibitemShut
  {NoStop}%
\bibitem [{\citenamefont {Troyer}\ and\ \citenamefont
  {Wiese}(2005)}]{Troyer:2004ge}%
  \BibitemOpen
  \bibfield  {author} {\bibinfo {author} {\bibfnamefont {M.}~\bibnamefont
  {Troyer}}\ and\ \bibinfo {author} {\bibfnamefont {U.-J.}\ \bibnamefont
  {Wiese}},\ }\bibfield  {title} {\bibinfo {title} {{Computational complexity
  and fundamental limitations to fermionic quantum Monte Carlo simulations}},\
  }\href {https://doi.org/10.1103/PhysRevLett.94.170201} {\bibfield  {journal}
  {\bibinfo  {journal} {Phys. Rev. Lett.}\ }\textbf {\bibinfo {volume} {94}},\
  \bibinfo {pages} {170201} (\bibinfo {year} {2005})},\ \Eprint
  {https://arxiv.org/abs/cond-mat/0408370} {arXiv:cond-mat/0408370}
  \BibitemShut {NoStop}%
\bibitem [{\citenamefont {Hiyama}\ \emph {et~al.}(2018)\citenamefont {Hiyama},
  \citenamefont {Hosaka}, \citenamefont {Oka},\ and\ \citenamefont
  {Richard}}]{Hiyama:2018ukv}%
  \BibitemOpen
  \bibfield  {author} {\bibinfo {author} {\bibfnamefont {E.}~\bibnamefont
  {Hiyama}}, \bibinfo {author} {\bibfnamefont {A.}~\bibnamefont {Hosaka}},
  \bibinfo {author} {\bibfnamefont {M.}~\bibnamefont {Oka}},\ and\ \bibinfo
  {author} {\bibfnamefont {J.-M.}\ \bibnamefont {Richard}},\ }\bibfield
  {title} {\bibinfo {title} {{Quark model estimate of hidden-charm pentaquark
  resonances}},\ }\href {https://doi.org/10.1103/PhysRevC.98.045208} {\bibfield
   {journal} {\bibinfo  {journal} {Phys. Rev. C}\ }\textbf {\bibinfo {volume}
  {98}},\ \bibinfo {pages} {045208} (\bibinfo {year} {2018})},\ \Eprint
  {https://arxiv.org/abs/1803.11369} {arXiv:1803.11369 [nucl-th]} \BibitemShut
  {NoStop}%
\bibitem [{\citenamefont {Giron}\ and\ \citenamefont
  {Lebed}(2021)}]{Giron:2021fnl}%
  \BibitemOpen
  \bibfield  {author} {\bibinfo {author} {\bibfnamefont {J.~F.}\ \bibnamefont
  {Giron}}\ and\ \bibinfo {author} {\bibfnamefont {R.~F.}\ \bibnamefont
  {Lebed}},\ }\bibfield  {title} {\bibinfo {title} {{Fine structure of
  pentaquark multiplets in the dynamical diquark model}},\ }\href
  {https://doi.org/10.1103/PhysRevD.104.114028} {\bibfield  {journal} {\bibinfo
   {journal} {Phys. Rev. D}\ }\textbf {\bibinfo {volume} {104}},\ \bibinfo
  {pages} {114028} (\bibinfo {year} {2021})},\ \Eprint
  {https://arxiv.org/abs/2110.05557} {arXiv:2110.05557 [hep-ph]} \BibitemShut
  {NoStop}%
\bibitem [{\citenamefont {Yan}\ \emph {et~al.}(2022)\citenamefont {Yan},
  \citenamefont {Wu}, \citenamefont {Hu}, \citenamefont {Huang},\ and\
  \citenamefont {Ping}}]{Yan:2021glh}%
  \BibitemOpen
  \bibfield  {author} {\bibinfo {author} {\bibfnamefont {Y.}~\bibnamefont
  {Yan}}, \bibinfo {author} {\bibfnamefont {Y.}~\bibnamefont {Wu}}, \bibinfo
  {author} {\bibfnamefont {X.}~\bibnamefont {Hu}}, \bibinfo {author}
  {\bibfnamefont {H.}~\bibnamefont {Huang}},\ and\ \bibinfo {author}
  {\bibfnamefont {J.}~\bibnamefont {Ping}},\ }\bibfield  {title} {\bibinfo
  {title} {{Fully heavy pentaquarks in quark models}},\ }\href
  {https://doi.org/10.1103/PhysRevD.105.014027} {\bibfield  {journal} {\bibinfo
   {journal} {Phys. Rev. D}\ }\textbf {\bibinfo {volume} {105}},\ \bibinfo
  {pages} {014027} (\bibinfo {year} {2022})},\ \Eprint
  {https://arxiv.org/abs/2110.10853} {arXiv:2110.10853 [hep-ph]} \BibitemShut
  {NoStop}%
\bibitem [{\citenamefont {Yang}\ \emph {et~al.}(2022)\citenamefont {Yang},
  \citenamefont {Ping},\ and\ \citenamefont {Segovia}}]{Yang:2022bfu}%
  \BibitemOpen
  \bibfield  {author} {\bibinfo {author} {\bibfnamefont {G.}~\bibnamefont
  {Yang}}, \bibinfo {author} {\bibfnamefont {J.}~\bibnamefont {Ping}},\ and\
  \bibinfo {author} {\bibfnamefont {J.}~\bibnamefont {Segovia}},\ }\bibfield
  {title} {\bibinfo {title} {{Fully charm and bottom pentaquarks in a
  lattice-QCD inspired quark model}},\ }\href
  {https://doi.org/10.1103/PhysRevD.106.014005} {\bibfield  {journal} {\bibinfo
   {journal} {Phys. Rev. D}\ }\textbf {\bibinfo {volume} {106}},\ \bibinfo
  {pages} {014005} (\bibinfo {year} {2022})},\ \Eprint
  {https://arxiv.org/abs/2205.11548} {arXiv:2205.11548 [hep-ph]} \BibitemShut
  {NoStop}%
\bibitem [{\citenamefont {An}\ \emph {et~al.}(2022)\citenamefont {An},
  \citenamefont {Luo}, \citenamefont {Liu},\ and\ \citenamefont
  {Liu}}]{An:2022fvs}%
  \BibitemOpen
  \bibfield  {author} {\bibinfo {author} {\bibfnamefont {H.-T.}\ \bibnamefont
  {An}}, \bibinfo {author} {\bibfnamefont {S.-Q.}\ \bibnamefont {Luo}},
  \bibinfo {author} {\bibfnamefont {Z.-W.}\ \bibnamefont {Liu}},\ and\ \bibinfo
  {author} {\bibfnamefont {X.}~\bibnamefont {Liu}},\ }\bibfield  {title}
  {\bibinfo {title} {{Fully heavy pentaquark states in constituent quark
  model}},\ }\href {https://doi.org/10.1103/PhysRevD.105.074032} {\bibfield
  {journal} {\bibinfo  {journal} {Phys. Rev. D}\ }\textbf {\bibinfo {volume}
  {105}},\ \bibinfo {pages} {074032} (\bibinfo {year} {2022})},\ \Eprint
  {https://arxiv.org/abs/2203.03448} {arXiv:2203.03448 [hep-ph]} \BibitemShut
  {NoStop}%
\bibitem [{\citenamefont {Liang}\ \emph {et~al.}(2025)\citenamefont {Liang},
  \citenamefont {Liu},\ and\ \citenamefont {Zhong}}]{Liang:2024met}%
  \BibitemOpen
  \bibfield  {author} {\bibinfo {author} {\bibfnamefont {Z.-B.}\ \bibnamefont
  {Liang}}, \bibinfo {author} {\bibfnamefont {F.-X.}\ \bibnamefont {Liu}},\
  and\ \bibinfo {author} {\bibfnamefont {X.-H.}\ \bibnamefont {Zhong}},\
  }\bibfield  {title} {\bibinfo {title} {{All-heavy pentaquarks}},\ }\href
  {https://doi.org/10.1103/PhysRevD.111.056013} {\bibfield  {journal} {\bibinfo
   {journal} {Phys. Rev. D}\ }\textbf {\bibinfo {volume} {111}},\ \bibinfo
  {pages} {056013} (\bibinfo {year} {2025})},\ \Eprint
  {https://arxiv.org/abs/2402.17974} {arXiv:2402.17974 [hep-ph]} \BibitemShut
  {NoStop}%
\bibitem [{\citenamefont {Gordillo}\ \emph {et~al.}(2024)\citenamefont
  {Gordillo}, \citenamefont {Segovia},\ and\ \citenamefont
  {Alcaraz-Pelegrina}}]{Gordillo:2024blx}%
  \BibitemOpen
  \bibfield  {author} {\bibinfo {author} {\bibfnamefont {M.~C.}\ \bibnamefont
  {Gordillo}}, \bibinfo {author} {\bibfnamefont {J.}~\bibnamefont {Segovia}},\
  and\ \bibinfo {author} {\bibfnamefont {J.~M.}\ \bibnamefont
  {Alcaraz-Pelegrina}},\ }\bibfield  {title} {\bibinfo {title} {{Diffusion
  Monte~Carlo calculation of fully heavy pentaquarks}},\ }\href
  {https://doi.org/10.1103/PhysRevD.110.094024} {\bibfield  {journal} {\bibinfo
   {journal} {Phys. Rev. D}\ }\textbf {\bibinfo {volume} {110}},\ \bibinfo
  {pages} {094024} (\bibinfo {year} {2024})},\ \Eprint
  {https://arxiv.org/abs/2409.04130} {arXiv:2409.04130 [hep-ph]} \BibitemShut
  {NoStop}%
\bibitem [{\citenamefont {LeCun}\ \emph {et~al.}(2015)\citenamefont {LeCun},
  \citenamefont {Bengio},\ and\ \citenamefont {Hinton}}]{LeCun:2015pmr}%
  \BibitemOpen
  \bibfield  {author} {\bibinfo {author} {\bibfnamefont {Y.}~\bibnamefont
  {LeCun}}, \bibinfo {author} {\bibfnamefont {Y.}~\bibnamefont {Bengio}},\ and\
  \bibinfo {author} {\bibfnamefont {G.}~\bibnamefont {Hinton}},\ }\bibfield
  {title} {\bibinfo {title} {{Deep learning}},\ }\href
  {https://doi.org/10.1038/nature14539} {\bibfield  {journal} {\bibinfo
  {journal} {Nature}\ }\textbf {\bibinfo {volume} {521}},\ \bibinfo {pages}
  {436} (\bibinfo {year} {2015})}\BibitemShut {NoStop}%
\bibitem [{\citenamefont {Carleo}\ \emph
  {et~al.}(2019{\natexlab{a}})\citenamefont {Carleo}, \citenamefont {Cirac},
  \citenamefont {Cranmer}, \citenamefont {Daudet}, \citenamefont {Schuld},
  \citenamefont {Tishby}, \citenamefont {Vogt-Maranto},\ and\ \citenamefont
  {Zdeborov\'a}}]{Carleo:2019ptp}%
  \BibitemOpen
  \bibfield  {author} {\bibinfo {author} {\bibfnamefont {G.}~\bibnamefont
  {Carleo}}, \bibinfo {author} {\bibfnamefont {I.}~\bibnamefont {Cirac}},
  \bibinfo {author} {\bibfnamefont {K.}~\bibnamefont {Cranmer}}, \bibinfo
  {author} {\bibfnamefont {L.}~\bibnamefont {Daudet}}, \bibinfo {author}
  {\bibfnamefont {M.}~\bibnamefont {Schuld}}, \bibinfo {author} {\bibfnamefont
  {N.}~\bibnamefont {Tishby}}, \bibinfo {author} {\bibfnamefont
  {L.}~\bibnamefont {Vogt-Maranto}},\ and\ \bibinfo {author} {\bibfnamefont
  {L.}~\bibnamefont {Zdeborov\'a}},\ }\bibfield  {title} {\bibinfo {title}
  {{Machine learning and the physical sciences}},\ }\href
  {https://doi.org/10.1103/RevModPhys.91.045002} {\bibfield  {journal}
  {\bibinfo  {journal} {Rev. Mod. Phys.}\ }\textbf {\bibinfo {volume} {91}},\
  \bibinfo {pages} {045002} (\bibinfo {year} {2019}{\natexlab{a}})},\ \Eprint
  {https://arxiv.org/abs/1903.10563} {arXiv:1903.10563 [physics.comp-ph]}
  \BibitemShut {NoStop}%
\bibitem [{\citenamefont {Zhang}\ \emph {et~al.}(2025)\citenamefont {Zhang},
  \citenamefont {Wang}, \citenamefont {Helwig}, \citenamefont {Luo},
  \citenamefont {Fu}, \citenamefont {Xie}, \citenamefont {Liu}, \citenamefont
  {Lin}, \citenamefont {Xu}, \citenamefont {Yan}, \citenamefont {Adams},
  \citenamefont {Weiler}, \citenamefont {Li}, \citenamefont {Fu}, \citenamefont
  {Wang}, \citenamefont {Strasser}, \citenamefont {Yu}, \citenamefont {Xie},
  \citenamefont {Fu}, \citenamefont {Xu}, \citenamefont {Liu}, \citenamefont
  {Du}, \citenamefont {Saxton}, \citenamefont {Ling}, \citenamefont {Lawrence},
  \citenamefont {Stärk}, \citenamefont {Gui}, \citenamefont {Edwards},
  \citenamefont {Gao}, \citenamefont {Ladera}, \citenamefont {Wu},
  \citenamefont {Hofgard}, \citenamefont {Tehrani}, \citenamefont {Wang},
  \citenamefont {Daigavane}, \citenamefont {Bohde}, \citenamefont {Kurtin},
  \citenamefont {Huang}, \citenamefont {Phung}, \citenamefont {Xu},
  \citenamefont {Joshi}, \citenamefont {Mathis}, \citenamefont
  {Azizzadenesheli}, \citenamefont {Fang}, \citenamefont {Aspuru-Guzik},
  \citenamefont {Bekkers}, \citenamefont {Bronstein}, \citenamefont {Zitnik},
  \citenamefont {Anandkumar}, \citenamefont {Ermon}, \citenamefont {Liò},
  \citenamefont {Yu}, \citenamefont {Günnemann}, \citenamefont {Leskovec},
  \citenamefont {Ji}, \citenamefont {Sun}, \citenamefont {Barzilay},
  \citenamefont {Jaakkola}, \citenamefont {Coley}, \citenamefont {Qian},
  \citenamefont {Qian}, \citenamefont {Smidt},\ and\ \citenamefont
  {Ji}}]{zhang2025AI4S}%
  \BibitemOpen
  \bibfield  {author} {\bibinfo {author} {\bibfnamefont {X.}~\bibnamefont
  {Zhang}}, \bibinfo {author} {\bibfnamefont {L.}~\bibnamefont {Wang}},
  \bibinfo {author} {\bibfnamefont {J.}~\bibnamefont {Helwig}}, \bibinfo
  {author} {\bibfnamefont {Y.}~\bibnamefont {Luo}}, \bibinfo {author}
  {\bibfnamefont {C.}~\bibnamefont {Fu}}, \bibinfo {author} {\bibfnamefont
  {Y.}~\bibnamefont {Xie}}, \bibinfo {author} {\bibfnamefont {M.}~\bibnamefont
  {Liu}}, \bibinfo {author} {\bibfnamefont {Y.}~\bibnamefont {Lin}}, \bibinfo
  {author} {\bibfnamefont {Z.}~\bibnamefont {Xu}}, \bibinfo {author}
  {\bibfnamefont {K.}~\bibnamefont {Yan}}, \bibinfo {author} {\bibfnamefont
  {K.}~\bibnamefont {Adams}}, \bibinfo {author} {\bibfnamefont
  {M.}~\bibnamefont {Weiler}}, \bibinfo {author} {\bibfnamefont
  {X.}~\bibnamefont {Li}}, \bibinfo {author} {\bibfnamefont {T.}~\bibnamefont
  {Fu}}, \bibinfo {author} {\bibfnamefont {Y.}~\bibnamefont {Wang}}, \bibinfo
  {author} {\bibfnamefont {A.}~\bibnamefont {Strasser}}, \bibinfo {author}
  {\bibfnamefont {H.}~\bibnamefont {Yu}}, \bibinfo {author} {\bibfnamefont
  {Y.}~\bibnamefont {Xie}}, \bibinfo {author} {\bibfnamefont {X.}~\bibnamefont
  {Fu}}, \bibinfo {author} {\bibfnamefont {S.}~\bibnamefont {Xu}}, \bibinfo
  {author} {\bibfnamefont {Y.}~\bibnamefont {Liu}}, \bibinfo {author}
  {\bibfnamefont {Y.}~\bibnamefont {Du}}, \bibinfo {author} {\bibfnamefont
  {A.}~\bibnamefont {Saxton}}, \bibinfo {author} {\bibfnamefont
  {H.}~\bibnamefont {Ling}}, \bibinfo {author} {\bibfnamefont {H.}~\bibnamefont
  {Lawrence}}, \bibinfo {author} {\bibfnamefont {H.}~\bibnamefont {Stärk}},
  \bibinfo {author} {\bibfnamefont {S.}~\bibnamefont {Gui}}, \bibinfo {author}
  {\bibfnamefont {C.}~\bibnamefont {Edwards}}, \bibinfo {author} {\bibfnamefont
  {N.}~\bibnamefont {Gao}}, \bibinfo {author} {\bibfnamefont {A.}~\bibnamefont
  {Ladera}}, \bibinfo {author} {\bibfnamefont {T.}~\bibnamefont {Wu}}, \bibinfo
  {author} {\bibfnamefont {E.~F.}\ \bibnamefont {Hofgard}}, \bibinfo {author}
  {\bibfnamefont {A.~M.}\ \bibnamefont {Tehrani}}, \bibinfo {author}
  {\bibfnamefont {R.}~\bibnamefont {Wang}}, \bibinfo {author} {\bibfnamefont
  {A.}~\bibnamefont {Daigavane}}, \bibinfo {author} {\bibfnamefont
  {M.}~\bibnamefont {Bohde}}, \bibinfo {author} {\bibfnamefont
  {J.}~\bibnamefont {Kurtin}}, \bibinfo {author} {\bibfnamefont
  {Q.}~\bibnamefont {Huang}}, \bibinfo {author} {\bibfnamefont
  {T.}~\bibnamefont {Phung}}, \bibinfo {author} {\bibfnamefont
  {M.}~\bibnamefont {Xu}}, \bibinfo {author} {\bibfnamefont {C.~K.}\
  \bibnamefont {Joshi}}, \bibinfo {author} {\bibfnamefont {S.~V.}\ \bibnamefont
  {Mathis}}, \bibinfo {author} {\bibfnamefont {K.}~\bibnamefont
  {Azizzadenesheli}}, \bibinfo {author} {\bibfnamefont {A.}~\bibnamefont
  {Fang}}, \bibinfo {author} {\bibfnamefont {A.}~\bibnamefont {Aspuru-Guzik}},
  \bibinfo {author} {\bibfnamefont {E.}~\bibnamefont {Bekkers}}, \bibinfo
  {author} {\bibfnamefont {M.}~\bibnamefont {Bronstein}}, \bibinfo {author}
  {\bibfnamefont {M.}~\bibnamefont {Zitnik}}, \bibinfo {author} {\bibfnamefont
  {A.}~\bibnamefont {Anandkumar}}, \bibinfo {author} {\bibfnamefont
  {S.}~\bibnamefont {Ermon}}, \bibinfo {author} {\bibfnamefont
  {P.}~\bibnamefont {Liò}}, \bibinfo {author} {\bibfnamefont {R.}~\bibnamefont
  {Yu}}, \bibinfo {author} {\bibfnamefont {S.}~\bibnamefont {Günnemann}},
  \bibinfo {author} {\bibfnamefont {J.}~\bibnamefont {Leskovec}}, \bibinfo
  {author} {\bibfnamefont {H.}~\bibnamefont {Ji}}, \bibinfo {author}
  {\bibfnamefont {J.}~\bibnamefont {Sun}}, \bibinfo {author} {\bibfnamefont
  {R.}~\bibnamefont {Barzilay}}, \bibinfo {author} {\bibfnamefont
  {T.}~\bibnamefont {Jaakkola}}, \bibinfo {author} {\bibfnamefont {C.~W.}\
  \bibnamefont {Coley}}, \bibinfo {author} {\bibfnamefont {X.}~\bibnamefont
  {Qian}}, \bibinfo {author} {\bibfnamefont {X.}~\bibnamefont {Qian}}, \bibinfo
  {author} {\bibfnamefont {T.}~\bibnamefont {Smidt}},\ and\ \bibinfo {author}
  {\bibfnamefont {S.}~\bibnamefont {Ji}},\ }\href
  {https://arxiv.org/abs/2307.08423} {\bibinfo {title} {Artificial intelligence
  for science in quantum, atomistic, and continuum systems}} (\bibinfo {year}
  {2025}),\ \Eprint {https://arxiv.org/abs/2307.08423} {arXiv:2307.08423
  [cs.LG]} \BibitemShut {NoStop}%
\bibitem [{\citenamefont {Carleo}\ and\ \citenamefont
  {Troyer}(2017)}]{Carleo2017}%
  \BibitemOpen
  \bibfield  {author} {\bibinfo {author} {\bibfnamefont {G.}~\bibnamefont
  {Carleo}}\ and\ \bibinfo {author} {\bibfnamefont {M.}~\bibnamefont
  {Troyer}},\ }\bibfield  {title} {\bibinfo {title} {{Solving the quantum
  many-body problem with artificial neural networks}},\ }\href
  {https://doi.org/10.1126/science.aag2302} {\bibfield  {journal} {\bibinfo
  {journal} {Science}\ }\textbf {\bibinfo {volume} {355}},\ \bibinfo {pages}
  {602} (\bibinfo {year} {2017})}\BibitemShut {NoStop}%
\bibitem [{\citenamefont {Han}\ \emph {et~al.}(2019)\citenamefont {Han},
  \citenamefont {Zhang},\ and\ \citenamefont {E}}]{Han2019}%
  \BibitemOpen
  \bibfield  {author} {\bibinfo {author} {\bibfnamefont {J.}~\bibnamefont
  {Han}}, \bibinfo {author} {\bibfnamefont {L.}~\bibnamefont {Zhang}},\ and\
  \bibinfo {author} {\bibfnamefont {W.}~\bibnamefont {E}},\ }\bibfield  {title}
  {\bibinfo {title} {Solving many-electron schrödinger equation using deep
  neural networks},\ }\href
  {https://doi.org/https://doi.org/10.1016/j.jcp.2019.108929} {\bibfield
  {journal} {\bibinfo  {journal} {Journal of Computational Physics}\ }\textbf
  {\bibinfo {volume} {399}},\ \bibinfo {pages} {108929} (\bibinfo {year}
  {2019})}\BibitemShut {NoStop}%
\bibitem [{\citenamefont {Pfau}\ \emph {et~al.}(2020)\citenamefont {Pfau},
  \citenamefont {Spencer}, \citenamefont {Matthews},\ and\ \citenamefont
  {Foulkes}}]{FermiNet}%
  \BibitemOpen
  \bibfield  {author} {\bibinfo {author} {\bibfnamefont {D.}~\bibnamefont
  {Pfau}}, \bibinfo {author} {\bibfnamefont {J.~S.}\ \bibnamefont {Spencer}},
  \bibinfo {author} {\bibfnamefont {A.~G. D.~G.}\ \bibnamefont {Matthews}},\
  and\ \bibinfo {author} {\bibfnamefont {W.~M.~C.}\ \bibnamefont {Foulkes}},\
  }\bibfield  {title} {\bibinfo {title} {Ab initio solution of the
  many-electron schr\"odinger equation with deep neural networks},\ }\href
  {https://doi.org/10.1103/PhysRevResearch.2.033429} {\bibfield  {journal}
  {\bibinfo  {journal} {Phys. Rev. Res.}\ }\textbf {\bibinfo {volume} {2}},\
  \bibinfo {pages} {033429} (\bibinfo {year} {2020})}\BibitemShut {NoStop}%
\bibitem [{\citenamefont {Hermann}\ \emph {et~al.}(2020)\citenamefont
  {Hermann}, \citenamefont {Sch{\"a}tzle},\ and\ \citenamefont
  {No{\'e}}}]{hermann2020deep}%
  \BibitemOpen
  \bibfield  {author} {\bibinfo {author} {\bibfnamefont {J.}~\bibnamefont
  {Hermann}}, \bibinfo {author} {\bibfnamefont {Z.}~\bibnamefont
  {Sch{\"a}tzle}},\ and\ \bibinfo {author} {\bibfnamefont {F.}~\bibnamefont
  {No{\'e}}},\ }\bibfield  {title} {\bibinfo {title} {Deep-neural-network
  solution of the electronic schr{\"o}dinger equation},\ }\href@noop {}
  {\bibfield  {journal} {\bibinfo  {journal} {Nature Chemistry}\ }\textbf
  {\bibinfo {volume} {12}},\ \bibinfo {pages} {891} (\bibinfo {year}
  {2020})}\BibitemShut {NoStop}%
\bibitem [{\citenamefont {Li}\ \emph {et~al.}(2022)\citenamefont {Li},
  \citenamefont {Li},\ and\ \citenamefont {Chen}}]{li2022ab}%
  \BibitemOpen
  \bibfield  {author} {\bibinfo {author} {\bibfnamefont {X.}~\bibnamefont
  {Li}}, \bibinfo {author} {\bibfnamefont {Z.}~\bibnamefont {Li}},\ and\
  \bibinfo {author} {\bibfnamefont {J.}~\bibnamefont {Chen}},\ }\bibfield
  {title} {\bibinfo {title} {Ab initio calculation of real solids via neural
  network ansatz},\ }\href@noop {} {\bibfield  {journal} {\bibinfo  {journal}
  {Nature Communications}\ }\textbf {\bibinfo {volume} {13}},\ \bibinfo {pages}
  {7895} (\bibinfo {year} {2022})}\BibitemShut {NoStop}%
\bibitem [{\citenamefont {Kim}\ \emph {et~al.}(2024)\citenamefont {Kim},
  \citenamefont {Pescia}, \citenamefont {Fore}, \citenamefont {Nys},
  \citenamefont {Carleo}, \citenamefont {Gandolfi}, \citenamefont
  {Hjorth-Jensen},\ and\ \citenamefont {Lovato}}]{Kim:2023fwy}%
  \BibitemOpen
  \bibfield  {author} {\bibinfo {author} {\bibfnamefont {J.}~\bibnamefont
  {Kim}}, \bibinfo {author} {\bibfnamefont {G.}~\bibnamefont {Pescia}},
  \bibinfo {author} {\bibfnamefont {B.}~\bibnamefont {Fore}}, \bibinfo {author}
  {\bibfnamefont {J.}~\bibnamefont {Nys}}, \bibinfo {author} {\bibfnamefont
  {G.}~\bibnamefont {Carleo}}, \bibinfo {author} {\bibfnamefont
  {S.}~\bibnamefont {Gandolfi}}, \bibinfo {author} {\bibfnamefont
  {M.}~\bibnamefont {Hjorth-Jensen}},\ and\ \bibinfo {author} {\bibfnamefont
  {A.}~\bibnamefont {Lovato}},\ }\bibfield  {title} {\bibinfo {title}
  {{Neural-network quantum states for ultra-cold Fermi gases}},\ }\href
  {https://doi.org/10.1038/s42005-024-01613-w} {\bibfield  {journal} {\bibinfo
  {journal} {Commun. Phys.}\ }\textbf {\bibinfo {volume} {7}},\ \bibinfo
  {pages} {148} (\bibinfo {year} {2024})},\ \Eprint
  {https://arxiv.org/abs/2305.08831} {arXiv:2305.08831 [cond-mat.quant-gas]}
  \BibitemShut {NoStop}%
\bibitem [{\citenamefont {Keeble}\ and\ \citenamefont
  {Rios}(2020)}]{Keeble2020}%
  \BibitemOpen
  \bibfield  {author} {\bibinfo {author} {\bibfnamefont {J.}~\bibnamefont
  {Keeble}}\ and\ \bibinfo {author} {\bibfnamefont {A.}~\bibnamefont {Rios}},\
  }\bibfield  {title} {\bibinfo {title} {Machine learning the deuteron},\
  }\href {https://doi.org/https://doi.org/10.1016/j.physletb.2020.135743}
  {\bibfield  {journal} {\bibinfo  {journal} {Physics Letters B}\ }\textbf
  {\bibinfo {volume} {809}},\ \bibinfo {pages} {135743} (\bibinfo {year}
  {2020})}\BibitemShut {NoStop}%
\bibitem [{\citenamefont {Adams}\ \emph {et~al.}(2021)\citenamefont {Adams},
  \citenamefont {Carleo}, \citenamefont {Lovato},\ and\ \citenamefont
  {Rocco}}]{Adams:2020aax}%
  \BibitemOpen
  \bibfield  {author} {\bibinfo {author} {\bibfnamefont {C.}~\bibnamefont
  {Adams}}, \bibinfo {author} {\bibfnamefont {G.}~\bibnamefont {Carleo}},
  \bibinfo {author} {\bibfnamefont {A.}~\bibnamefont {Lovato}},\ and\ \bibinfo
  {author} {\bibfnamefont {N.}~\bibnamefont {Rocco}},\ }\bibfield  {title}
  {\bibinfo {title} {{Variational Monte Carlo Calculations of
  A\ensuremath{\leq}4 Nuclei with an Artificial Neural-Network Correlator
  Ansatz}},\ }\href {https://doi.org/10.1103/PhysRevLett.127.022502} {\bibfield
   {journal} {\bibinfo  {journal} {Phys. Rev. Lett.}\ }\textbf {\bibinfo
  {volume} {127}},\ \bibinfo {pages} {022502} (\bibinfo {year} {2021})},\
  \Eprint {https://arxiv.org/abs/2007.14282} {arXiv:2007.14282 [nucl-th]}
  \BibitemShut {NoStop}%
\bibitem [{\citenamefont {Yang}\ and\ \citenamefont
  {Zhao}(2022)}]{Yang:2022esu}%
  \BibitemOpen
  \bibfield  {author} {\bibinfo {author} {\bibfnamefont {Y.~L.}\ \bibnamefont
  {Yang}}\ and\ \bibinfo {author} {\bibfnamefont {P.~W.}\ \bibnamefont
  {Zhao}},\ }\bibfield  {title} {\bibinfo {title} {{A consistent description of
  the relativistic effects and three-body interactions in atomic nuclei}},\
  }\href {https://doi.org/10.1016/j.physletb.2022.137587} {\bibfield  {journal}
  {\bibinfo  {journal} {Phys. Lett. B}\ }\textbf {\bibinfo {volume} {835}},\
  \bibinfo {pages} {137587} (\bibinfo {year} {2022})},\ \Eprint
  {https://arxiv.org/abs/2206.13208} {arXiv:2206.13208 [nucl-th]} \BibitemShut
  {NoStop}%
\bibitem [{\citenamefont {Yang}\ and\ \citenamefont
  {Zhao}(2023)}]{Yang:2022rlw}%
  \BibitemOpen
  \bibfield  {author} {\bibinfo {author} {\bibfnamefont {Y.}~\bibnamefont
  {Yang}}\ and\ \bibinfo {author} {\bibfnamefont {P.}~\bibnamefont {Zhao}},\
  }\bibfield  {title} {\bibinfo {title} {{Deep-neural-network approach to
  solving the ab initio nuclear structure problem}},\ }\href
  {https://doi.org/10.1103/PhysRevC.107.034320} {\bibfield  {journal} {\bibinfo
   {journal} {Phys. Rev. C}\ }\textbf {\bibinfo {volume} {107}},\ \bibinfo
  {pages} {034320} (\bibinfo {year} {2023})},\ \Eprint
  {https://arxiv.org/abs/2211.13998} {arXiv:2211.13998 [nucl-th]} \BibitemShut
  {NoStop}%
\bibitem [{\citenamefont {Yang}\ and\ \citenamefont
  {Zhao}(2025)}]{Yang:2024wsg}%
  \BibitemOpen
  \bibfield  {author} {\bibinfo {author} {\bibfnamefont {Y.-L.}\ \bibnamefont
  {Yang}}\ and\ \bibinfo {author} {\bibfnamefont {P.-W.}\ \bibnamefont
  {Zhao}},\ }\bibfield  {title} {\bibinfo {title} {{Reconciling Light Nuclei
  and Nuclear Matter: Relativistic ab initio Calculations}},\ }\href
  {https://doi.org/10.1088/0256-307X/42/5/051201} {\bibfield  {journal}
  {\bibinfo  {journal} {Chin. Phys. Lett.}\ }\textbf {\bibinfo {volume} {42}},\
  \bibinfo {pages} {051201} (\bibinfo {year} {2025})},\ \Eprint
  {https://arxiv.org/abs/2405.04203} {arXiv:2405.04203 [nucl-th]} \BibitemShut
  {NoStop}%
\bibitem [{\citenamefont {Fore}\ \emph {et~al.}(2025)\citenamefont {Fore},
  \citenamefont {Kim}, \citenamefont {Hjorth-Jensen},\ and\ \citenamefont
  {Lovato}}]{Fore:2024exa}%
  \BibitemOpen
  \bibfield  {author} {\bibinfo {author} {\bibfnamefont {B.}~\bibnamefont
  {Fore}}, \bibinfo {author} {\bibfnamefont {J.}~\bibnamefont {Kim}}, \bibinfo
  {author} {\bibfnamefont {M.}~\bibnamefont {Hjorth-Jensen}},\ and\ \bibinfo
  {author} {\bibfnamefont {A.}~\bibnamefont {Lovato}},\ }\bibfield  {title}
  {\bibinfo {title} {{Investigating the crust of neutron stars with
  neural-network quantum states}},\ }\href
  {https://doi.org/10.1038/s42005-025-02015-2} {\bibfield  {journal} {\bibinfo
  {journal} {Commun. Phys.}\ }\textbf {\bibinfo {volume} {8}},\ \bibinfo
  {pages} {108} (\bibinfo {year} {2025})},\ \Eprint
  {https://arxiv.org/abs/2407.21207} {arXiv:2407.21207 [nucl-th]} \BibitemShut
  {NoStop}%
\bibitem [{\citenamefont {Yang}\ \emph {et~al.}(2025)\citenamefont {Yang},
  \citenamefont {Epelbaum}, \citenamefont {Meng}, \citenamefont {Meng},\ and\
  \citenamefont {Zhao}}]{Yang:2025mhg}%
  \BibitemOpen
  \bibfield  {author} {\bibinfo {author} {\bibfnamefont {Y.}~\bibnamefont
  {Yang}}, \bibinfo {author} {\bibfnamefont {E.}~\bibnamefont {Epelbaum}},
  \bibinfo {author} {\bibfnamefont {J.}~\bibnamefont {Meng}}, \bibinfo {author}
  {\bibfnamefont {L.}~\bibnamefont {Meng}},\ and\ \bibinfo {author}
  {\bibfnamefont {P.}~\bibnamefont {Zhao}},\ }\bibfield  {title} {\bibinfo
  {title} {{Chiral symmetry and peripheral neutron-$\alpha$ scattering}},\
  }\href@noop {} {\  (\bibinfo {year} {2025})},\ \Eprint
  {https://arxiv.org/abs/2502.09961} {arXiv:2502.09961 [nucl-th]} \BibitemShut
  {NoStop}%
\bibitem [{\citenamefont {Hornik}\ \emph {et~al.}(1989)\citenamefont {Hornik},
  \citenamefont {Stinchcombe},\ and\ \citenamefont {White}}]{HORNIK1989359}%
  \BibitemOpen
  \bibfield  {author} {\bibinfo {author} {\bibfnamefont {K.}~\bibnamefont
  {Hornik}}, \bibinfo {author} {\bibfnamefont {M.}~\bibnamefont
  {Stinchcombe}},\ and\ \bibinfo {author} {\bibfnamefont {H.}~\bibnamefont
  {White}},\ }\bibfield  {title} {\bibinfo {title} {Multilayer feedforward
  networks are universal approximators},\ }\href
  {https://doi.org/https://doi.org/10.1016/0893-6080(89)90020-8} {\bibfield
  {journal} {\bibinfo  {journal} {Neural Networks}\ }\textbf {\bibinfo {volume}
  {2}},\ \bibinfo {pages} {359} (\bibinfo {year} {1989})}\BibitemShut {NoStop}%
\bibitem [{\citenamefont {Kidger}\ and\ \citenamefont
  {Lyons}(2020)}]{kidger2020}%
  \BibitemOpen
  \bibfield  {author} {\bibinfo {author} {\bibfnamefont {P.}~\bibnamefont
  {Kidger}}\ and\ \bibinfo {author} {\bibfnamefont {T.}~\bibnamefont {Lyons}},\
  }\href {https://arxiv.org/abs/1905.08539} {\bibinfo {title} {Universal
  approximation with deep narrow networks}} (\bibinfo {year} {2020}),\ \Eprint
  {https://arxiv.org/abs/1905.08539} {arXiv:1905.08539 [cs.LG]} \BibitemShut
  {NoStop}%
\bibitem [{\citenamefont {Semay}\ and\ \citenamefont
  {Silvestre-Brac}(1994)}]{Semay:1994ht}%
  \BibitemOpen
  \bibfield  {author} {\bibinfo {author} {\bibfnamefont {C.}~\bibnamefont
  {Semay}}\ and\ \bibinfo {author} {\bibfnamefont {B.}~\bibnamefont
  {Silvestre-Brac}},\ }\bibfield  {title} {\bibinfo {title} {{Diquonia and
  potential models}},\ }\href {https://doi.org/10.1007/BF01413104} {\bibfield
  {journal} {\bibinfo  {journal} {Z. Phys. C}\ }\textbf {\bibinfo {volume}
  {61}},\ \bibinfo {pages} {271} (\bibinfo {year} {1994})}\BibitemShut
  {NoStop}%
\bibitem [{\citenamefont {Silvestre-Brac}(1996)}]{SilvestreBrac1996}%
  \BibitemOpen
  \bibfield  {author} {\bibinfo {author} {\bibfnamefont {B.}~\bibnamefont
  {Silvestre-Brac}},\ }\bibfield  {title} {\bibinfo {title} {{Spectrum and
  static properties of heavy baryons}},\ }\href
  {https://doi.org/10.1007/s006010050028} {\bibfield  {journal} {\bibinfo
  {journal} {Few Body Syst.}\ }\textbf {\bibinfo {volume} {20}},\ \bibinfo
  {pages} {1} (\bibinfo {year} {1996})}\BibitemShut {NoStop}%
\bibitem [{\citenamefont {Eichten}\ \emph {et~al.}(1978)\citenamefont
  {Eichten}, \citenamefont {Gottfried}, \citenamefont {Kinoshita},
  \citenamefont {Lane},\ and\ \citenamefont {Yan}}]{Eichten:1978tg}%
  \BibitemOpen
  \bibfield  {author} {\bibinfo {author} {\bibfnamefont {E.}~\bibnamefont
  {Eichten}}, \bibinfo {author} {\bibfnamefont {K.}~\bibnamefont {Gottfried}},
  \bibinfo {author} {\bibfnamefont {T.}~\bibnamefont {Kinoshita}}, \bibinfo
  {author} {\bibfnamefont {K.~D.}\ \bibnamefont {Lane}},\ and\ \bibinfo
  {author} {\bibfnamefont {T.-M.}\ \bibnamefont {Yan}},\ }\bibfield  {title}
  {\bibinfo {title} {{Charmonium: The Model}},\ }\href
  {https://doi.org/10.1103/PhysRevD.17.3090} {\bibfield  {journal} {\bibinfo
  {journal} {Phys. Rev. D}\ }\textbf {\bibinfo {volume} {17}},\ \bibinfo
  {pages} {3090} (\bibinfo {year} {1978})},\ \bibinfo {note} {[Erratum:
  Phys.Rev.D 21, 313 (1980)]}\BibitemShut {NoStop}%
\bibitem [{\citenamefont {Eichten}\ \emph {et~al.}(1980)\citenamefont
  {Eichten}, \citenamefont {Gottfried}, \citenamefont {Kinoshita},
  \citenamefont {Lane},\ and\ \citenamefont {Yan}}]{Eichten:1979ms}%
  \BibitemOpen
  \bibfield  {author} {\bibinfo {author} {\bibfnamefont {E.}~\bibnamefont
  {Eichten}}, \bibinfo {author} {\bibfnamefont {K.}~\bibnamefont {Gottfried}},
  \bibinfo {author} {\bibfnamefont {T.}~\bibnamefont {Kinoshita}}, \bibinfo
  {author} {\bibfnamefont {K.~D.}\ \bibnamefont {Lane}},\ and\ \bibinfo
  {author} {\bibfnamefont {T.-M.}\ \bibnamefont {Yan}},\ }\bibfield  {title}
  {\bibinfo {title} {{Charmonium: Comparison with Experiment}},\ }\href
  {https://doi.org/10.1103/PhysRevD.21.203} {\bibfield  {journal} {\bibinfo
  {journal} {Phys. Rev. D}\ }\textbf {\bibinfo {volume} {21}},\ \bibinfo
  {pages} {203} (\bibinfo {year} {1980})}\BibitemShut {NoStop}%
\bibitem [{\citenamefont {Artru}(1975)}]{Artru:1974zn}%
  \BibitemOpen
  \bibfield  {author} {\bibinfo {author} {\bibfnamefont {X.}~\bibnamefont
  {Artru}},\ }\bibfield  {title} {\bibinfo {title} {{String Model with Baryons:
  Topology, Classical Motion}},\ }\href
  {https://doi.org/10.1016/0550-3213(75)90019-X} {\bibfield  {journal}
  {\bibinfo  {journal} {Nucl. Phys. B}\ }\textbf {\bibinfo {volume} {85}},\
  \bibinfo {pages} {442} (\bibinfo {year} {1975})}\BibitemShut {NoStop}%
\bibitem [{\citenamefont {Takahashi}\ \emph {et~al.}(2001)\citenamefont
  {Takahashi}, \citenamefont {Matsufuru}, \citenamefont {Nemoto},\ and\
  \citenamefont {Suganuma}}]{Takahashi:2000te}%
  \BibitemOpen
  \bibfield  {author} {\bibinfo {author} {\bibfnamefont {T.~T.}\ \bibnamefont
  {Takahashi}}, \bibinfo {author} {\bibfnamefont {H.}~\bibnamefont
  {Matsufuru}}, \bibinfo {author} {\bibfnamefont {Y.}~\bibnamefont {Nemoto}},\
  and\ \bibinfo {author} {\bibfnamefont {H.}~\bibnamefont {Suganuma}},\
  }\bibfield  {title} {\bibinfo {title} {{The Three quark potential in the
  SU(3) lattice QCD}},\ }\href {https://doi.org/10.1103/PhysRevLett.86.18}
  {\bibfield  {journal} {\bibinfo  {journal} {Phys. Rev. Lett.}\ }\textbf
  {\bibinfo {volume} {86}},\ \bibinfo {pages} {18} (\bibinfo {year} {2001})},\
  \Eprint {https://arxiv.org/abs/hep-lat/0006005} {arXiv:hep-lat/0006005}
  \BibitemShut {NoStop}%
\bibitem [{\citenamefont {Takahashi}\ \emph {et~al.}(2002)\citenamefont
  {Takahashi}, \citenamefont {Suganuma}, \citenamefont {Nemoto},\ and\
  \citenamefont {Matsufuru}}]{Takahashi:2002bw}%
  \BibitemOpen
  \bibfield  {author} {\bibinfo {author} {\bibfnamefont {T.~T.}\ \bibnamefont
  {Takahashi}}, \bibinfo {author} {\bibfnamefont {H.}~\bibnamefont {Suganuma}},
  \bibinfo {author} {\bibfnamefont {Y.}~\bibnamefont {Nemoto}},\ and\ \bibinfo
  {author} {\bibfnamefont {H.}~\bibnamefont {Matsufuru}},\ }\bibfield  {title}
  {\bibinfo {title} {{Detailed analysis of the three quark potential in SU(3)
  lattice QCD}},\ }\href {https://doi.org/10.1103/PhysRevD.65.114509}
  {\bibfield  {journal} {\bibinfo  {journal} {Phys. Rev. D}\ }\textbf {\bibinfo
  {volume} {65}},\ \bibinfo {pages} {114509} (\bibinfo {year} {2002})},\
  \Eprint {https://arxiv.org/abs/hep-lat/0204011} {arXiv:hep-lat/0204011}
  \BibitemShut {NoStop}%
\bibitem [{Sup()}]{Suppl}%
  \BibitemOpen
  \href@noop {} {}\bibinfo {note} {{See Supplemental Material for additional
  details on the quark potential model, neural-network optimization process and
  results of electron systems., which include
  Refs.~\cite{Semay:1994ht,SilvestreBrac1996,Ma:2022vqf,Vijande:2004he,
  Yang:2022rlw, Takahashi:2002bw, Kinghorn:1993zz, ParticleDataGroup:2024cfk,
  Segovia:2011dg, Meng:2023jqk, Takahashi:2000te, Ho:1993zz,
  Adams:2020aax,Hastings1970,Metropolis1953,PhysRevB.71.241103,netket3:2022,YiBoYangQuarkMassLow2023,PhysRevA.74.052502}}}\BibitemShut
  {NoStop}%
\bibitem [{\citenamefont {Ma}\ \emph {et~al.}(2025)\citenamefont {Ma},
  \citenamefont {Meng}, \citenamefont {Wen},\ and\ \citenamefont
  {Zhu}}]{Ma:2025rvj}%
  \BibitemOpen
  \bibfield  {author} {\bibinfo {author} {\bibfnamefont {Y.}~\bibnamefont
  {Ma}}, \bibinfo {author} {\bibfnamefont {L.}~\bibnamefont {Meng}}, \bibinfo
  {author} {\bibfnamefont {L.-Z.}\ \bibnamefont {Wen}},\ and\ \bibinfo {author}
  {\bibfnamefont {S.-L.}\ \bibnamefont {Zhu}},\ }\bibfield  {title} {\bibinfo
  {title} {{Trilepton and tetralepton bound and resonant states: The QED
  counterpart of multiquark states}},\ }\href
  {https://doi.org/10.1103/PhysRevD.111.073001} {\bibfield  {journal} {\bibinfo
   {journal} {Phys. Rev. D}\ }\textbf {\bibinfo {volume} {111}},\ \bibinfo
  {pages} {073001} (\bibinfo {year} {2025})},\ \Eprint
  {https://arxiv.org/abs/2501.00871} {arXiv:2501.00871 [hep-ph]} \BibitemShut
  {NoStop}%
\bibitem [{\citenamefont {Ma}\ \emph {et~al.}(2023)\citenamefont {Ma},
  \citenamefont {Meng}, \citenamefont {Chen},\ and\ \citenamefont
  {Zhu}}]{Ma:2022vqf}%
  \BibitemOpen
  \bibfield  {author} {\bibinfo {author} {\bibfnamefont {Y.}~\bibnamefont
  {Ma}}, \bibinfo {author} {\bibfnamefont {L.}~\bibnamefont {Meng}}, \bibinfo
  {author} {\bibfnamefont {Y.-K.}\ \bibnamefont {Chen}},\ and\ \bibinfo
  {author} {\bibfnamefont {S.-L.}\ \bibnamefont {Zhu}},\ }\bibfield  {title}
  {\bibinfo {title} {{Ground state baryons in the flux-tube three-body
  confinement model using diffusion Monte~Carlo}},\ }\href
  {https://doi.org/10.1103/PhysRevD.107.054035} {\bibfield  {journal} {\bibinfo
   {journal} {Phys. Rev. D}\ }\textbf {\bibinfo {volume} {107}},\ \bibinfo
  {pages} {054035} (\bibinfo {year} {2023})},\ \Eprint
  {https://arxiv.org/abs/2211.09021} {arXiv:2211.09021 [hep-ph]} \BibitemShut
  {NoStop}%
\bibitem [{\citenamefont {Meng}\ \emph
  {et~al.}(2023{\natexlab{b}})\citenamefont {Meng}, \citenamefont {Chen},
  \citenamefont {Ma},\ and\ \citenamefont {Zhu}}]{Meng:2023jqk}%
  \BibitemOpen
  \bibfield  {author} {\bibinfo {author} {\bibfnamefont {L.}~\bibnamefont
  {Meng}}, \bibinfo {author} {\bibfnamefont {Y.-K.}\ \bibnamefont {Chen}},
  \bibinfo {author} {\bibfnamefont {Y.}~\bibnamefont {Ma}},\ and\ \bibinfo
  {author} {\bibfnamefont {S.-L.}\ \bibnamefont {Zhu}},\ }\bibfield  {title}
  {\bibinfo {title} {{Tetraquark bound states in constituent quark models:
  Benchmark test calculations}},\ }\href
  {https://doi.org/10.1103/PhysRevD.108.114016} {\bibfield  {journal} {\bibinfo
   {journal} {Phys. Rev. D}\ }\textbf {\bibinfo {volume} {108}},\ \bibinfo
  {pages} {114016} (\bibinfo {year} {2023}{\natexlab{b}})},\ \Eprint
  {https://arxiv.org/abs/2310.13354} {arXiv:2310.13354 [hep-ph]} \BibitemShut
  {NoStop}%
\bibitem [{\citenamefont {Zouzou}\ \emph {et~al.}(1986)\citenamefont {Zouzou},
  \citenamefont {Silvestre-Brac}, \citenamefont {Gignoux},\ and\ \citenamefont
  {Richard}}]{Zouzou1986}%
  \BibitemOpen
  \bibfield  {author} {\bibinfo {author} {\bibfnamefont {S.}~\bibnamefont
  {Zouzou}}, \bibinfo {author} {\bibfnamefont {B.}~\bibnamefont
  {Silvestre-Brac}}, \bibinfo {author} {\bibfnamefont {C.}~\bibnamefont
  {Gignoux}},\ and\ \bibinfo {author} {\bibfnamefont {J.~M.}\ \bibnamefont
  {Richard}},\ }\bibfield  {title} {\bibinfo {title} {{FOUR QUARK BOUND
  STATES}},\ }\href {https://doi.org/10.1007/BF01557611} {\bibfield  {journal}
  {\bibinfo  {journal} {Z. Phys. C}\ }\textbf {\bibinfo {volume} {30}},\
  \bibinfo {pages} {457} (\bibinfo {year} {1986})}\BibitemShut {NoStop}%
\bibitem [{\citenamefont {Manohar}\ and\ \citenamefont
  {Wise}(1993)}]{Manohar:1992nd}%
  \BibitemOpen
  \bibfield  {author} {\bibinfo {author} {\bibfnamefont {A.~V.}\ \bibnamefont
  {Manohar}}\ and\ \bibinfo {author} {\bibfnamefont {M.~B.}\ \bibnamefont
  {Wise}},\ }\bibfield  {title} {\bibinfo {title} {{Exotic Q Q anti-q anti-q
  states in QCD}},\ }\href {https://doi.org/10.1016/0550-3213(93)90614-U}
  {\bibfield  {journal} {\bibinfo  {journal} {Nucl. Phys. B}\ }\textbf
  {\bibinfo {volume} {399}},\ \bibinfo {pages} {17} (\bibinfo {year} {1993})},\
  \Eprint {https://arxiv.org/abs/hep-ph/9212236} {arXiv:hep-ph/9212236}
  \BibitemShut {NoStop}%
\bibitem [{\citenamefont {Francis}\ \emph {et~al.}(2017)\citenamefont
  {Francis}, \citenamefont {Hudspith}, \citenamefont {Lewis},\ and\
  \citenamefont {Maltman}}]{Francis:2016hui}%
  \BibitemOpen
  \bibfield  {author} {\bibinfo {author} {\bibfnamefont {A.}~\bibnamefont
  {Francis}}, \bibinfo {author} {\bibfnamefont {R.~J.}\ \bibnamefont
  {Hudspith}}, \bibinfo {author} {\bibfnamefont {R.}~\bibnamefont {Lewis}},\
  and\ \bibinfo {author} {\bibfnamefont {K.}~\bibnamefont {Maltman}},\
  }\bibfield  {title} {\bibinfo {title} {{Lattice Prediction for Deeply Bound
  Doubly Heavy Tetraquarks}},\ }\href
  {https://doi.org/10.1103/PhysRevLett.118.142001} {\bibfield  {journal}
  {\bibinfo  {journal} {Phys. Rev. Lett.}\ }\textbf {\bibinfo {volume} {118}},\
  \bibinfo {pages} {142001} (\bibinfo {year} {2017})},\ \Eprint
  {https://arxiv.org/abs/1607.05214} {arXiv:1607.05214 [hep-lat]} \BibitemShut
  {NoStop}%
\bibitem [{\citenamefont {Junnarkar}\ \emph {et~al.}(2019)\citenamefont
  {Junnarkar}, \citenamefont {Mathur},\ and\ \citenamefont
  {Padmanath}}]{Junnarkar:2018twb}%
  \BibitemOpen
  \bibfield  {author} {\bibinfo {author} {\bibfnamefont {P.}~\bibnamefont
  {Junnarkar}}, \bibinfo {author} {\bibfnamefont {N.}~\bibnamefont {Mathur}},\
  and\ \bibinfo {author} {\bibfnamefont {M.}~\bibnamefont {Padmanath}},\
  }\bibfield  {title} {\bibinfo {title} {{Study of doubly heavy tetraquarks in
  Lattice QCD}},\ }\href {https://doi.org/10.1103/PhysRevD.99.034507}
  {\bibfield  {journal} {\bibinfo  {journal} {Phys. Rev. D}\ }\textbf {\bibinfo
  {volume} {99}},\ \bibinfo {pages} {034507} (\bibinfo {year} {2019})},\
  \Eprint {https://arxiv.org/abs/1810.12285} {arXiv:1810.12285 [hep-lat]}
  \BibitemShut {NoStop}%
\bibitem [{\citenamefont {Hayrapetyan}\ \emph {et~al.}(2024)\citenamefont
  {Hayrapetyan} \emph {et~al.}}]{CMS:2023owd}%
  \BibitemOpen
  \bibfield  {author} {\bibinfo {author} {\bibfnamefont {A.}~\bibnamefont
  {Hayrapetyan}} \emph {et~al.} (\bibinfo {collaboration} {CMS}),\ }\bibfield
  {title} {\bibinfo {title} {{New Structures in the
  J/\ensuremath{\psi}J/\ensuremath{\psi} Mass Spectrum in Proton-Proton
  Collisions at s=13\,\,TeV}},\ }\href
  {https://doi.org/10.1103/PhysRevLett.132.111901} {\bibfield  {journal}
  {\bibinfo  {journal} {Phys. Rev. Lett.}\ }\textbf {\bibinfo {volume} {132}},\
  \bibinfo {pages} {111901} (\bibinfo {year} {2024})},\ \Eprint
  {https://arxiv.org/abs/2306.07164} {arXiv:2306.07164 [hep-ex]} \BibitemShut
  {NoStop}%
\bibitem [{\citenamefont {Aad}\ \emph {et~al.}(2023)\citenamefont {Aad} \emph
  {et~al.}}]{ATLAS:2023bft}%
  \BibitemOpen
  \bibfield  {author} {\bibinfo {author} {\bibfnamefont {G.}~\bibnamefont
  {Aad}} \emph {et~al.} (\bibinfo {collaboration} {ATLAS}),\ }\bibfield
  {title} {\bibinfo {title} {{Observation of an Excess of Dicharmonium Events
  in the Four-Muon Final State with the ATLAS Detector}},\ }\href
  {https://doi.org/10.1103/PhysRevLett.131.151902} {\bibfield  {journal}
  {\bibinfo  {journal} {Phys. Rev. Lett.}\ }\textbf {\bibinfo {volume} {131}},\
  \bibinfo {pages} {151902} (\bibinfo {year} {2023})},\ \Eprint
  {https://arxiv.org/abs/2304.08962} {arXiv:2304.08962 [hep-ex]} \BibitemShut
  {NoStop}%
\bibitem [{\citenamefont {Hayrapetyan}\ \emph {et~al.}(2025)\citenamefont
  {Hayrapetyan} \emph {et~al.}}]{CMS:2025fpt}%
  \BibitemOpen
  \bibfield  {author} {\bibinfo {author} {\bibfnamefont {A.}~\bibnamefont
  {Hayrapetyan}} \emph {et~al.} (\bibinfo {collaboration} {CMS}),\ }\bibfield
  {title} {\bibinfo {title} {{Determination of the spin and parity of all-charm
  tetraquarks}},\ }\href {https://doi.org/10.1038/s41586-025-09711-7}
  {\bibfield  {journal} {\bibinfo  {journal} {Nature}\ }\textbf {\bibinfo
  {volume} {648}},\ \bibinfo {pages} {58} (\bibinfo {year} {2025})},\ \Eprint
  {https://arxiv.org/abs/2506.07944} {arXiv:2506.07944 [hep-ex]} \BibitemShut
  {NoStop}%
\bibitem [{\citenamefont {Wu}\ \emph {et~al.}(2024{\natexlab{a}})\citenamefont
  {Wu}, \citenamefont {Chen}, \citenamefont {Meng},\ and\ \citenamefont
  {Zhu}}]{Wu:2024euj}%
  \BibitemOpen
  \bibfield  {author} {\bibinfo {author} {\bibfnamefont {W.-L.}\ \bibnamefont
  {Wu}}, \bibinfo {author} {\bibfnamefont {Y.-K.}\ \bibnamefont {Chen}},
  \bibinfo {author} {\bibfnamefont {L.}~\bibnamefont {Meng}},\ and\ \bibinfo
  {author} {\bibfnamefont {S.-L.}\ \bibnamefont {Zhu}},\ }\bibfield  {title}
  {\bibinfo {title} {{Benchmark calculations of fully heavy compact and
  molecular tetraquark states}},\ }\href
  {https://doi.org/10.1103/PhysRevD.109.054034} {\bibfield  {journal} {\bibinfo
   {journal} {Phys. Rev. D}\ }\textbf {\bibinfo {volume} {109}},\ \bibinfo
  {pages} {054034} (\bibinfo {year} {2024}{\natexlab{a}})},\ \Eprint
  {https://arxiv.org/abs/2401.14899} {arXiv:2401.14899 [hep-ph]} \BibitemShut
  {NoStop}%
\bibitem [{\citenamefont {Li}\ \emph {et~al.}(2013)\citenamefont {Li},
  \citenamefont {Sun}, \citenamefont {Liu},\ and\ \citenamefont
  {Zhu}}]{Li:2012ss}%
  \BibitemOpen
  \bibfield  {author} {\bibinfo {author} {\bibfnamefont {N.}~\bibnamefont
  {Li}}, \bibinfo {author} {\bibfnamefont {Z.-F.}\ \bibnamefont {Sun}},
  \bibinfo {author} {\bibfnamefont {X.}~\bibnamefont {Liu}},\ and\ \bibinfo
  {author} {\bibfnamefont {S.-L.}\ \bibnamefont {Zhu}},\ }\bibfield  {title}
  {\bibinfo {title} {{Coupled-channel analysis of the possible $D^{(*)}D^{(*)},
  \overline{B}^{(*)}\overline{B}^{(*)}$ and $D^{(*)}\overline{B}^{(*)}$
  molecular states}},\ }\href {https://doi.org/10.1103/PhysRevD.88.114008}
  {\bibfield  {journal} {\bibinfo  {journal} {Phys. Rev. D}\ }\textbf {\bibinfo
  {volume} {88}},\ \bibinfo {pages} {114008} (\bibinfo {year} {2013})},\
  \Eprint {https://arxiv.org/abs/1211.5007} {arXiv:1211.5007 [hep-ph]}
  \BibitemShut {NoStop}%
\bibitem [{\citenamefont {Wu}\ \emph {et~al.}(2024{\natexlab{b}})\citenamefont
  {Wu}, \citenamefont {Ma}, \citenamefont {Chen}, \citenamefont {Meng},\ and\
  \citenamefont {Zhu}}]{Wu:2024zbx}%
  \BibitemOpen
  \bibfield  {author} {\bibinfo {author} {\bibfnamefont {W.-L.}\ \bibnamefont
  {Wu}}, \bibinfo {author} {\bibfnamefont {Y.}~\bibnamefont {Ma}}, \bibinfo
  {author} {\bibfnamefont {Y.-K.}\ \bibnamefont {Chen}}, \bibinfo {author}
  {\bibfnamefont {L.}~\bibnamefont {Meng}},\ and\ \bibinfo {author}
  {\bibfnamefont {S.-L.}\ \bibnamefont {Zhu}},\ }\bibfield  {title} {\bibinfo
  {title} {{Doubly heavy tetraquark bound and resonant states}},\ }\href
  {https://doi.org/10.1103/PhysRevD.110.094041} {\bibfield  {journal} {\bibinfo
   {journal} {Phys. Rev. D}\ }\textbf {\bibinfo {volume} {110}},\ \bibinfo
  {pages} {094041} (\bibinfo {year} {2024}{\natexlab{b}})},\ \Eprint
  {https://arxiv.org/abs/2409.03373} {arXiv:2409.03373 [hep-ph]} \BibitemShut
  {NoStop}%
\bibitem [{\citenamefont {Guo}\ \emph {et~al.}(2013)\citenamefont {Guo},
  \citenamefont {Hidalgo-Duque}, \citenamefont {Nieves},\ and\ \citenamefont
  {Valderrama}}]{Guo:2013xga}%
  \BibitemOpen
  \bibfield  {author} {\bibinfo {author} {\bibfnamefont {F.-K.}\ \bibnamefont
  {Guo}}, \bibinfo {author} {\bibfnamefont {C.}~\bibnamefont {Hidalgo-Duque}},
  \bibinfo {author} {\bibfnamefont {J.}~\bibnamefont {Nieves}},\ and\ \bibinfo
  {author} {\bibfnamefont {M.~P.}\ \bibnamefont {Valderrama}},\ }\bibfield
  {title} {\bibinfo {title} {{Heavy-antiquark\textendash{}diquark symmetry and
  heavy hadron molecules: Are there triply heavy pentaquarks?}},\ }\href
  {https://doi.org/10.1103/PhysRevD.88.054014} {\bibfield  {journal} {\bibinfo
  {journal} {Phys. Rev. D}\ }\textbf {\bibinfo {volume} {88}},\ \bibinfo
  {pages} {054014} (\bibinfo {year} {2013})},\ \Eprint
  {https://arxiv.org/abs/1305.4052} {arXiv:1305.4052 [hep-ph]} \BibitemShut
  {NoStop}%
\bibitem [{\citenamefont {Wang}\ \emph {et~al.}(2024)\citenamefont {Wang},
  \citenamefont {Xiao}, \citenamefont {Sun},\ and\ \citenamefont
  {Liu}}]{Wang:2024yjp}%
  \BibitemOpen
  \bibfield  {author} {\bibinfo {author} {\bibfnamefont {Z.-Y.}\ \bibnamefont
  {Wang}}, \bibinfo {author} {\bibfnamefont {C.-W.}\ \bibnamefont {Xiao}},
  \bibinfo {author} {\bibfnamefont {Z.-F.}\ \bibnamefont {Sun}},\ and\ \bibinfo
  {author} {\bibfnamefont {X.}~\bibnamefont {Liu}},\ }\bibfield  {title}
  {\bibinfo {title} {{Possible molecules of triple-heavy pentaquarks within the
  extended local hidden gauge formalism}},\ }\href
  {https://doi.org/10.1103/PhysRevD.110.076014} {\bibfield  {journal} {\bibinfo
   {journal} {Phys. Rev. D}\ }\textbf {\bibinfo {volume} {110}},\ \bibinfo
  {pages} {076014} (\bibinfo {year} {2024})},\ \Eprint
  {https://arxiv.org/abs/2407.13319} {arXiv:2407.13319 [hep-ph]} \BibitemShut
  {NoStop}%
\bibitem [{\citenamefont {Bradbury}\ \emph {et~al.}(2018)\citenamefont
  {Bradbury}, \citenamefont {Frostig}, \citenamefont {Hawkins}, \citenamefont
  {Johnson}, \citenamefont {Leary}, \citenamefont {Maclaurin}, \citenamefont
  {Necula}, \citenamefont {Paszke}, \citenamefont {VanderPlas}, \citenamefont
  {{Wanderman-Milne}},\ and\ \citenamefont {Zhang}}]{jax2018github}%
  \BibitemOpen
  \bibfield  {author} {\bibinfo {author} {\bibfnamefont {J.}~\bibnamefont
  {Bradbury}}, \bibinfo {author} {\bibfnamefont {R.}~\bibnamefont {Frostig}},
  \bibinfo {author} {\bibfnamefont {P.}~\bibnamefont {Hawkins}}, \bibinfo
  {author} {\bibfnamefont {M.~J.}\ \bibnamefont {Johnson}}, \bibinfo {author}
  {\bibfnamefont {C.}~\bibnamefont {Leary}}, \bibinfo {author} {\bibfnamefont
  {D.}~\bibnamefont {Maclaurin}}, \bibinfo {author} {\bibfnamefont
  {G.}~\bibnamefont {Necula}}, \bibinfo {author} {\bibfnamefont
  {A.}~\bibnamefont {Paszke}}, \bibinfo {author} {\bibfnamefont
  {J.}~\bibnamefont {VanderPlas}}, \bibinfo {author} {\bibfnamefont
  {S.}~\bibnamefont {{Wanderman-Milne}}},\ and\ \bibinfo {author}
  {\bibfnamefont {Q.}~\bibnamefont {Zhang}},\ }\href@noop {} {\bibinfo {title}
  {{{JAX}}: Composable transformations of {{Python}}+{{NumPy}} programs}}
  (\bibinfo {year} {2018})\BibitemShut {NoStop}%
\bibitem [{\citenamefont {Heek}\ \emph {et~al.}(2024)\citenamefont {Heek},
  \citenamefont {Levskaya}, \citenamefont {Oliver}, \citenamefont {Ritter},
  \citenamefont {Rondepierre}, \citenamefont {Steiner},\ and\ \citenamefont
  {{van Zee}}}]{flax2020github}%
  \BibitemOpen
  \bibfield  {author} {\bibinfo {author} {\bibfnamefont {J.}~\bibnamefont
  {Heek}}, \bibinfo {author} {\bibfnamefont {A.}~\bibnamefont {Levskaya}},
  \bibinfo {author} {\bibfnamefont {A.}~\bibnamefont {Oliver}}, \bibinfo
  {author} {\bibfnamefont {M.}~\bibnamefont {Ritter}}, \bibinfo {author}
  {\bibfnamefont {B.}~\bibnamefont {Rondepierre}}, \bibinfo {author}
  {\bibfnamefont {A.}~\bibnamefont {Steiner}},\ and\ \bibinfo {author}
  {\bibfnamefont {M.}~\bibnamefont {{van Zee}}},\ }\href@noop {} {\bibinfo
  {title} {Flax: {{A}} neural network library and ecosystem for {{JAX}}}}
  (\bibinfo {year} {2024})\BibitemShut {NoStop}%
\bibitem [{\citenamefont {Vicentini}\ \emph {et~al.}(2022)\citenamefont
  {Vicentini}, \citenamefont {Hofmann}, \citenamefont {Szabó}, \citenamefont
  {Wu}, \citenamefont {Roth}, \citenamefont {Giuliani}, \citenamefont {Pescia},
  \citenamefont {Nys}, \citenamefont {Vargas-Calderón}, \citenamefont
  {Astrakhantsev},\ and\ \citenamefont {Carleo}}]{netket3:2022}%
  \BibitemOpen
  \bibfield  {author} {\bibinfo {author} {\bibfnamefont {F.}~\bibnamefont
  {Vicentini}}, \bibinfo {author} {\bibfnamefont {D.}~\bibnamefont {Hofmann}},
  \bibinfo {author} {\bibfnamefont {A.}~\bibnamefont {Szabó}}, \bibinfo
  {author} {\bibfnamefont {D.}~\bibnamefont {Wu}}, \bibinfo {author}
  {\bibfnamefont {C.}~\bibnamefont {Roth}}, \bibinfo {author} {\bibfnamefont
  {C.}~\bibnamefont {Giuliani}}, \bibinfo {author} {\bibfnamefont
  {G.}~\bibnamefont {Pescia}}, \bibinfo {author} {\bibfnamefont
  {J.}~\bibnamefont {Nys}}, \bibinfo {author} {\bibfnamefont {V.}~\bibnamefont
  {Vargas-Calderón}}, \bibinfo {author} {\bibfnamefont {N.}~\bibnamefont
  {Astrakhantsev}},\ and\ \bibinfo {author} {\bibfnamefont {G.}~\bibnamefont
  {Carleo}},\ }\bibfield  {title} {\bibinfo {title} {Netket 3: Machine learning
  toolbox for many-body quantum systems},\ }\href
  {https://doi.org/10.21468/SciPostPhysCodeb.7} {\bibfield  {journal} {\bibinfo
   {journal} {SciPost Phys. Codebases}\ ,\ \bibinfo {pages} {7}} (\bibinfo
  {year} {2022})}\BibitemShut {NoStop}%
\bibitem [{\citenamefont {Carleo}\ \emph
  {et~al.}(2019{\natexlab{b}})\citenamefont {Carleo}, \citenamefont {Choo},
  \citenamefont {Hofmann}, \citenamefont {Smith}, \citenamefont {Westerhout},
  \citenamefont {Alet}, \citenamefont {Davis}, \citenamefont {Efthymiou},
  \citenamefont {Glasser}, \citenamefont {Lin}, \citenamefont {Mauri},
  \citenamefont {Mazzola}, \citenamefont {Pereira},\ and\ \citenamefont
  {Vicentini}}]{netket2:2019}%
  \BibitemOpen
  \bibfield  {author} {\bibinfo {author} {\bibfnamefont {G.}~\bibnamefont
  {Carleo}}, \bibinfo {author} {\bibfnamefont {K.}~\bibnamefont {Choo}},
  \bibinfo {author} {\bibfnamefont {D.}~\bibnamefont {Hofmann}}, \bibinfo
  {author} {\bibfnamefont {J.~E.}\ \bibnamefont {Smith}}, \bibinfo {author}
  {\bibfnamefont {T.}~\bibnamefont {Westerhout}}, \bibinfo {author}
  {\bibfnamefont {F.}~\bibnamefont {Alet}}, \bibinfo {author} {\bibfnamefont
  {E.~J.}\ \bibnamefont {Davis}}, \bibinfo {author} {\bibfnamefont
  {S.}~\bibnamefont {Efthymiou}}, \bibinfo {author} {\bibfnamefont
  {I.}~\bibnamefont {Glasser}}, \bibinfo {author} {\bibfnamefont {S.-H.}\
  \bibnamefont {Lin}}, \bibinfo {author} {\bibfnamefont {M.}~\bibnamefont
  {Mauri}}, \bibinfo {author} {\bibfnamefont {G.}~\bibnamefont {Mazzola}},
  \bibinfo {author} {\bibfnamefont {C.~B.}\ \bibnamefont {Pereira}},\ and\
  \bibinfo {author} {\bibfnamefont {F.}~\bibnamefont {Vicentini}},\ }\bibfield
  {title} {\bibinfo {title} {Netket: A machine learning toolkit for many-body
  quantum systems},\ }\href {https://doi.org/10.1016/j.softx.2019.100311}
  {\bibfield  {journal} {\bibinfo  {journal} {SoftwareX}\ }\textbf {\bibinfo
  {volume} {10}},\ \bibinfo {pages} {100311} (\bibinfo {year}
  {2019}{\natexlab{b}})}\BibitemShut {NoStop}%
\bibitem [{\citenamefont {Wu}(2026)}]{Wu_DeepQuark_2026}%
  \BibitemOpen
  \bibfield  {author} {\bibinfo {author} {\bibfnamefont {W.-L.}\ \bibnamefont
  {Wu}},\ }\href {https://github.com/wlwuphy/DeepQuark} {\bibinfo {title}
  {{DeepQuark}}} (\bibinfo {year} {2026})\BibitemShut {NoStop}%
\bibitem [{\citenamefont {Vijande}\ \emph {et~al.}(2005)\citenamefont
  {Vijande}, \citenamefont {Fernandez},\ and\ \citenamefont
  {Valcarce}}]{Vijande:2004he}%
  \BibitemOpen
  \bibfield  {author} {\bibinfo {author} {\bibfnamefont {J.}~\bibnamefont
  {Vijande}}, \bibinfo {author} {\bibfnamefont {F.}~\bibnamefont {Fernandez}},\
  and\ \bibinfo {author} {\bibfnamefont {A.}~\bibnamefont {Valcarce}},\
  }\bibfield  {title} {\bibinfo {title} {{Constituent quark model study of the
  meson spectra}},\ }\href {https://doi.org/10.1088/0954-3899/31/5/017}
  {\bibfield  {journal} {\bibinfo  {journal} {J. Phys. G}\ }\textbf {\bibinfo
  {volume} {31}},\ \bibinfo {pages} {481} (\bibinfo {year} {2005})},\ \Eprint
  {https://arxiv.org/abs/hep-ph/0411299} {arXiv:hep-ph/0411299} \BibitemShut
  {NoStop}%
\bibitem [{\citenamefont {Kinghorn}\ and\ \citenamefont
  {Poshusta}(1993)}]{Kinghorn:1993zz}%
  \BibitemOpen
  \bibfield  {author} {\bibinfo {author} {\bibfnamefont {D.~B.}\ \bibnamefont
  {Kinghorn}}\ and\ \bibinfo {author} {\bibfnamefont {R.~D.}\ \bibnamefont
  {Poshusta}},\ }\bibfield  {title} {\bibinfo {title} {{Nonadiabatic
  variational calculations on dipositronium using explicitly correlated
  Gaussian basis functions}},\ }\href
  {https://doi.org/10.1103/PhysRevA.47.3671} {\bibfield  {journal} {\bibinfo
  {journal} {Phys. Rev. A}\ }\textbf {\bibinfo {volume} {47}},\ \bibinfo
  {pages} {3671} (\bibinfo {year} {1993})}\BibitemShut {NoStop}%
\bibitem [{\citenamefont {Navas}\ \emph {et~al.}(2024)\citenamefont {Navas}
  \emph {et~al.}}]{ParticleDataGroup:2024cfk}%
  \BibitemOpen
  \bibfield  {author} {\bibinfo {author} {\bibfnamefont {S.}~\bibnamefont
  {Navas}} \emph {et~al.} (\bibinfo {collaboration} {Particle Data Group}),\
  }\bibfield  {title} {\bibinfo {title} {{Review of particle physics}},\ }\href
  {https://doi.org/10.1103/PhysRevD.110.030001} {\bibfield  {journal} {\bibinfo
   {journal} {Phys. Rev. D}\ }\textbf {\bibinfo {volume} {110}},\ \bibinfo
  {pages} {030001} (\bibinfo {year} {2024})}\BibitemShut {NoStop}%
\bibitem [{\citenamefont {Segovia}\ \emph {et~al.}(2011)\citenamefont
  {Segovia}, \citenamefont {Albertus}, \citenamefont {Entem}, \citenamefont
  {Fernandez}, \citenamefont {Hernandez},\ and\ \citenamefont
  {Perez-Garcia}}]{Segovia:2011dg}%
  \BibitemOpen
  \bibfield  {author} {\bibinfo {author} {\bibfnamefont {J.}~\bibnamefont
  {Segovia}}, \bibinfo {author} {\bibfnamefont {C.}~\bibnamefont {Albertus}},
  \bibinfo {author} {\bibfnamefont {D.~R.}\ \bibnamefont {Entem}}, \bibinfo
  {author} {\bibfnamefont {F.}~\bibnamefont {Fernandez}}, \bibinfo {author}
  {\bibfnamefont {E.}~\bibnamefont {Hernandez}},\ and\ \bibinfo {author}
  {\bibfnamefont {M.~A.}\ \bibnamefont {Perez-Garcia}},\ }\bibfield  {title}
  {\bibinfo {title} {{Semileptonic $B$ and $B_{s}$ decays into orbitally
  excited charmed mesons}},\ }\href
  {https://doi.org/10.1103/PhysRevD.84.094029} {\bibfield  {journal} {\bibinfo
  {journal} {Phys. Rev. D}\ }\textbf {\bibinfo {volume} {84}},\ \bibinfo
  {pages} {094029} (\bibinfo {year} {2011})},\ \Eprint
  {https://arxiv.org/abs/1107.4248} {arXiv:1107.4248 [hep-ph]} \BibitemShut
  {NoStop}%
\bibitem [{\citenamefont {Ho}(1993)}]{Ho:1993zz}%
  \BibitemOpen
  \bibfield  {author} {\bibinfo {author} {\bibfnamefont {Y.~K.}\ \bibnamefont
  {Ho}},\ }\bibfield  {title} {\bibinfo {title} {{Variational calculation of
  ground-state energy of positronium negative ions}},\ }\href
  {https://doi.org/10.1103/PhysRevA.48.4780} {\bibfield  {journal} {\bibinfo
  {journal} {Phys. Rev. A}\ }\textbf {\bibinfo {volume} {48}},\ \bibinfo
  {pages} {4780} (\bibinfo {year} {1993})}\BibitemShut {NoStop}%
\bibitem [{\citenamefont {Hastings}(1970)}]{Hastings1970}%
  \BibitemOpen
  \bibfield  {author} {\bibinfo {author} {\bibfnamefont {W.~K.}\ \bibnamefont
  {Hastings}},\ }\bibfield  {title} {\bibinfo {title} {Monte carlo sampling
  methods using markov chains and their applications},\ }\href
  {https://doi.org/10.1093/biomet/57.1.97} {\bibfield  {journal} {\bibinfo
  {journal} {Biometrika}\ }\textbf {\bibinfo {volume} {57}},\ \bibinfo {pages}
  {97} (\bibinfo {year} {1970})},\ \Eprint
  {https://arxiv.org/abs/https://academic.oup.com/biomet/article-pdf/57/1/97/23940249/57-1-97.pdf}
  {https://academic.oup.com/biomet/article-pdf/57/1/97/23940249/57-1-97.pdf}
  \BibitemShut {NoStop}%
\bibitem [{\citenamefont {Metropolis}\ \emph {et~al.}(1953)\citenamefont
  {Metropolis}, \citenamefont {Rosenbluth}, \citenamefont {Rosenbluth},
  \citenamefont {Teller},\ and\ \citenamefont {Teller}}]{Metropolis1953}%
  \BibitemOpen
  \bibfield  {author} {\bibinfo {author} {\bibfnamefont {N.}~\bibnamefont
  {Metropolis}}, \bibinfo {author} {\bibfnamefont {A.~W.}\ \bibnamefont
  {Rosenbluth}}, \bibinfo {author} {\bibfnamefont {M.~N.}\ \bibnamefont
  {Rosenbluth}}, \bibinfo {author} {\bibfnamefont {A.~H.}\ \bibnamefont
  {Teller}},\ and\ \bibinfo {author} {\bibfnamefont {E.}~\bibnamefont
  {Teller}},\ }\bibfield  {title} {\bibinfo {title} {Equation of state
  calculations by fast computing machines},\ }\href
  {https://doi.org/10.1063/1.1699114} {\bibfield  {journal} {\bibinfo
  {journal} {The Journal of Chemical Physics}\ }\textbf {\bibinfo {volume}
  {21}},\ \bibinfo {pages} {1087} (\bibinfo {year} {1953})},\ \Eprint
  {https://arxiv.org/abs/https://pubs.aip.org/aip/jcp/article-pdf/21/6/1087/18802390/1087\_1\_online.pdf}
  {https://pubs.aip.org/aip/jcp/article-pdf/21/6/1087/18802390/1087\_1\_online.pdf}
  \BibitemShut {NoStop}%
\bibitem [{\citenamefont {Sorella}(2005)}]{PhysRevB.71.241103}%
  \BibitemOpen
  \bibfield  {author} {\bibinfo {author} {\bibfnamefont {S.}~\bibnamefont
  {Sorella}},\ }\bibfield  {title} {\bibinfo {title} {Wave function
  optimization in the variational monte carlo method},\ }\href
  {https://doi.org/10.1103/PhysRevB.71.241103} {\bibfield  {journal} {\bibinfo
  {journal} {Phys. Rev. B}\ }\textbf {\bibinfo {volume} {71}},\ \bibinfo
  {pages} {241103} (\bibinfo {year} {2005})}\BibitemShut {NoStop}%
\bibitem [{\citenamefont {Yang}(2023)}]{YiBoYangQuarkMassLow2023}%
  \BibitemOpen
  \bibfield  {author} {\bibinfo {author} {\bibfnamefont {Y.-B.}\ \bibnamefont
  {Yang}},\ }\bibfield  {title} {\bibinfo {title} {Quark mass and low energy
  constant in the continuum using the {{CLQCD}} ensembles},\ }\href
  {https://indico.ihep.ac.cn/event/19002/contributions/142210/} {\bibfield
  {journal} {\bibinfo  {journal} {The 3th Chinese Lattice QCD workshop}\ }
  (\bibinfo {year} {2023})}\BibitemShut {NoStop}%
\bibitem [{\citenamefont {Bubin}\ and\ \citenamefont
  {Adamowicz}(2006)}]{PhysRevA.74.052502}%
  \BibitemOpen
  \bibfield  {author} {\bibinfo {author} {\bibfnamefont {S.}~\bibnamefont
  {Bubin}}\ and\ \bibinfo {author} {\bibfnamefont {L.}~\bibnamefont
  {Adamowicz}},\ }\bibfield  {title} {\bibinfo {title} {Nonrelativistic
  variational calculations of the positronium molecule and the positronium
  hydride},\ }\href {https://doi.org/10.1103/PhysRevA.74.052502} {\bibfield
  {journal} {\bibinfo  {journal} {Phys. Rev. A}\ }\textbf {\bibinfo {volume}
  {74}},\ \bibinfo {pages} {052502} (\bibinfo {year} {2006})}\BibitemShut
  {NoStop}%
\end{thebibliography}%

\clearpage
\onecolumngrid
   \setcounter{equation}{0}
\setcounter{figure}{0}
\setcounter{table}{0}
\setcounter{page}{1}
\makeatletter
\renewcommand{\theequation}{S\arabic{equation}}
\renewcommand{\thefigure}{S\arabic{figure}}
\renewcommand\thetable{SM-\Roman{table}}  
\renewcommand{\appendixname}{Supplemental Material}


\appendix

\begin{center}
\textbf{\large Supplemental Materials: DeepQuark: deep-neural-network approach to multiquark bound states}
\end{center}
\begin{mdframed}[hidealllines=true,innerleftmargin=0.1\textwidth,innerrightmargin=0.1\textwidth]
~~This supplemental material provides additional details on the quark potential model, neural-network optimization process and results of electron systems.
\end{mdframed}


\section{Quark Potential Models}\label{SM:QM}
\subsection{The AL1 potential and flux-tube potential}
In a nonrelativisitic quark potential model, the Hamiltonian of a multiquark system in the center-of-mass frame reads,
\begin{equation}
	H=\sum_{i=1}^n\left(m_i+\frac{p_i^2}{2m_i}\right)+\sum_{i<j=1}^nV_{ij}.
\end{equation}
In this work, we adopt the AL1 potential~\cite{Semay:1994ht,SilvestreBrac1996}, consisting of the one-gluon-exchange (OGE) interaction and two-body linear confinement interaction,
\begin{equation}
	\label{eqsm:AL1}
	\begin{aligned}
		&V_{\text{OGE},ij} =-\frac{3}{16} \boldsymbol\lambda_i \cdot \boldsymbol\lambda_j\left(-\frac{\kappa}{r_{i j}}-\Lambda+\frac{8 \pi \kappa^{\prime}}{3 m_i m_j} \frac{e^{ -r_{i j}^2 / r_0^2}}{\pi^{3 / 2} r_0^3} \boldsymbol{s}_i \cdot \boldsymbol{s}_j\right),\\
		&V_{\text{conf},ij}=-\frac{3}{16} \boldsymbol\lambda_i \cdot \boldsymbol\lambda_j \lambda r_{i j}.
	\end{aligned}
\end{equation}
For baryon systems, an additional phenomenological term $V_{123}=-\frac{C}{m_1m_2m_3}$ is introduced to mimic the three-body interaction effect~\cite{SilvestreBrac1996}. For hadron systems with heavy quarks, this term is suppressed by the heavy quark mass and therefore is neglected in the calculations. The parameters of the AL1 potential are taken from Ref.~\cite{SilvestreBrac1996} and listed in Table~\ref{tabsm:paraAL1}. The masses of heavy mesons and baryons in the model are given in Table~\ref{tabsm:meson_baryon}.
\begin{table*}[htbp]
	\centering
	\caption{The parameters in the AL1 quark potential model.}
	\label{tabsm:paraAL1}
	\begin{tabular*}{\hsize}{@{}@{\extracolsep{\fill}}ccccccccccc@{}}
		\hline\hline
		$ \kappa $ &$ \lambda { [\mathrm{GeV}^{2}]}$&$ \Lambda {\rm [GeV]} $&$ \kappa^\prime $&$ m_b {\rm [GeV]}$&$ m_c {\rm [GeV]}$&$ m_q {\rm [GeV]}$& $r_0 {\rm [GeV^{-1}]}$ &$ A { [\mathrm{GeV}^{B-1}]}$&$ B $&$C [\mathrm{GeV}^4]$\\
		\hline
		0.5069&0.1653&0.8321&1.8609&5.227&1.836&0.315&$A\left(\frac{2m_im_j}{m_i+m_j}\right)^{-B}$&1.6553&0.2204&2.02$\times 10^{-3}$\\
		\hline\hline
	\end{tabular*}
\end{table*}
\begin{table*}
	\centering
	\caption{The masses (in MeV) of heavy mesons and baryons in the AL1 quark potential model. The results from Gaussian expansion method and the experimental values are shown for comparison. }
	\label{tabsm:meson_baryon}
	\begin{tabular*}{\hsize}{@{}@{\extracolsep{\fill}}lcccc@{}}
		\hline\hline
		&$J^P$& DQ&GEM~\cite{Ma:2022vqf,Meng:2023jqk}&EXP~\cite{ParticleDataGroup:2024cfk} \\
		\hline
		$D(c\bar q)$&$0^-$&1862&1862&1867\\
		$\bar B(b\bar q)$&&5294&5294&5279\\
		$\eta_c(c\bar c)$ & &3005&3005&2984\\
		$\eta_b(b\bar b)$ & &9424&9424&9399\\
		$D^*(c\bar q)$&$1^-$&2016&2016&2009\\
		$\bar B^*(b\bar q)$&&5350&5350&5325\\
		$J/\psi(c\bar c)$ & &3101&3101&3097 \\
		$\Upsilon(b\bar b)$ & &9462&9462&9460 \\
		$\Lambda_c^+(cqq)$&$\frac{1}{2}^+$&2290&2291&2286\\
		$\Lambda_b^0(bqq)$&&5636&5636&5620\\
		$\Xi_{cc}^*(ccq)$&$\frac{3}{2}^+$&3702&3702&$\cdots$\\
		$\Xi_{bb}^*(bbq)$&&10232&10232&$\cdots$\\
		\hline\hline
	\end{tabular*}
\end{table*}
\begin{figure}[h]
	\centering
	\includegraphics[width=0.3\textwidth]{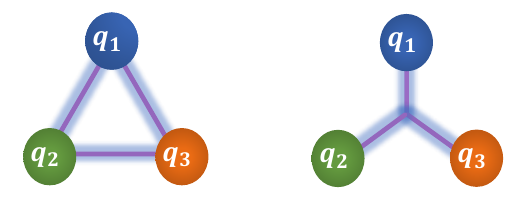}
	\caption{Two confinement scenarios for the baryons. The left and right panels represent the pairwise confinement mechanism  and the flux-tube confinement mechanism, respectively.}
	\label{figsm:qqqFT}
\end{figure}
For the nucleon system, we also test the flux-tube confinement potential as shown in Fig.~\ref{figsm:qqqFT}, where the linear pairwise confinement interactions $V_{\text{conf},ij}$ are replaced by the flux-tube interaction $V^{\text{ft}}_{\text{conf}}=\sigma L_{\text{min}}$. The string tension $\sigma$ is given as $\sigma=0.9204\lambda$~\cite{Ma:2022vqf}, which was determined by fitting the experimental mass of $\Omega^-(sss)$ and is consistent with the best fitting parameters in lattice QCD simulation~\cite{Takahashi:2000te}. In baryon systems, the minimal length of color flux tubes can be solved analytically~\cite{Takahashi:2002bw} and is given as 

	\begin{equation}
		L_{\text{min}}=\begin{cases}
			\left[\frac{1}{2}(a^2+b^2+c^2)+\frac{\sqrt{3}}{2}\sqrt{d(d-2a)(d-2b)(d-2c)}\right]^{1/2}&\max(\theta_a,\theta_b,\theta_c)<\frac{2\pi}{3},\\
			d-\max(a,b,c)&\max(\theta_a,\theta_b,\theta_c)>\frac{2\pi}{3},
		\end{cases}
	\end{equation}
where $\theta_a,\theta_b,\theta_c$ are the interior angles of the baryon triangles,  $a,b,c$ are the lengths of sides, and $d=a+b+c$.

\subsection{Uncertainty of the AL1 model}~\label{app:al1error}
Generally, the uncertainty of a phenomenological model consists of two parts: the discrepancy between the model and the underlying theory and the uncertainty of the model parameters.

{\it The discrepancy between the model and the underlying theory:} The AL1 model include some core ingredients of QCD, such as the short-range one-gluon-exchange interaction and the long-range confinement. However, given the non-perturbative nature of QCD in low-energy regime, there are still many unknown factors in the quark-quark interactions, such as the effects of loop corrections. These effects may be partly compensated by fitting the model parameters to reproduce the experimental meson spectra.  However, some factors may be missing in the model, and it is highly challenging to quantify such uncertainties, as it would require full knowledge of the non-perturbative QCD. We can only infer from the meson spectra that the energy results of the quark potential model may deviate from the experimental values by around tens of MeV.
	
{\it The uncertainty of the model parameters:} The parameters in the AL1 model include four constituent quark masses and six potential parameters. They were determined by fitting the meson spectra across all flavor sectors~\cite{SilvestreBrac1996}. Given the high dimensionality of the parameter space and the strong correlations among parameters, a comprehensive determination of parameter uncertainties would require an extensive exploration of the full ten-dimensional parameter space, which is highly computationally demanding. Instead, we employ a sampling-and-selection strategy as a practical approach to provide a reasonable estimate of parameter uncertainties. We vary the parameters within a range around their best-fit values, such that the goodness of fit of the meson spectra remains comparable. 
	
\begin{figure}[ht]
	\centering
	\includegraphics[width=\textwidth]{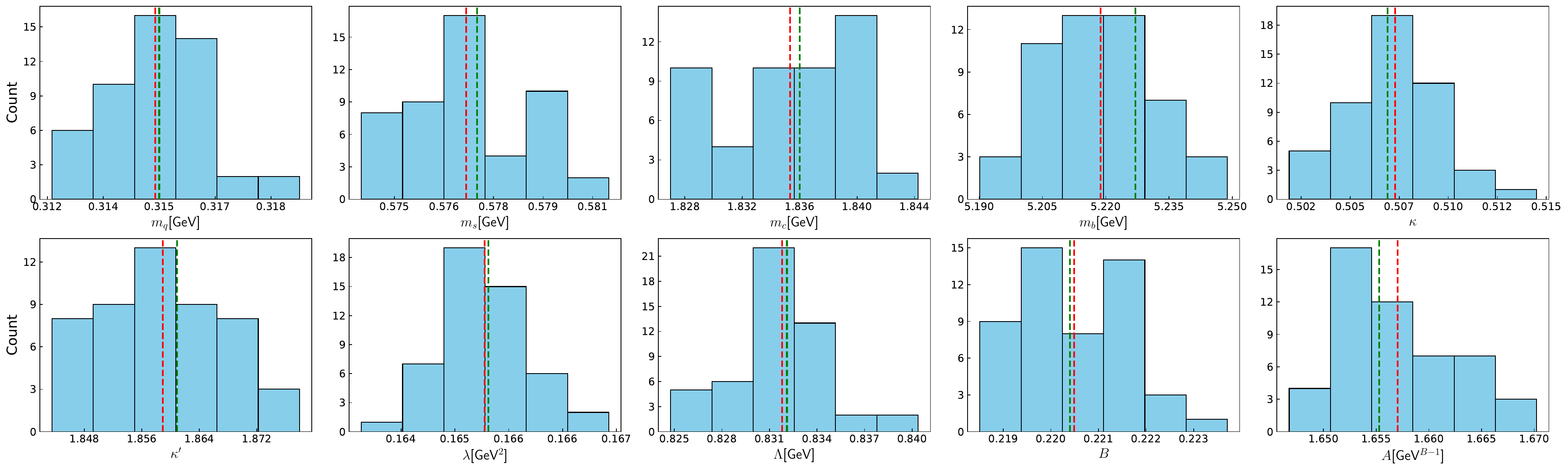}
	\caption{Distributions of the AL1 model parameter values in the 50 selected parameter sets. The red and green dashed lines represent the mean values over the 50 parameter sets and the original values in Ref.~\cite{SilvestreBrac1996}, respectively.}
	\label{figsm:para_dis}
\end{figure}

\begin{figure}[ht]
	\centering
	\includegraphics[width=0.8\textwidth]{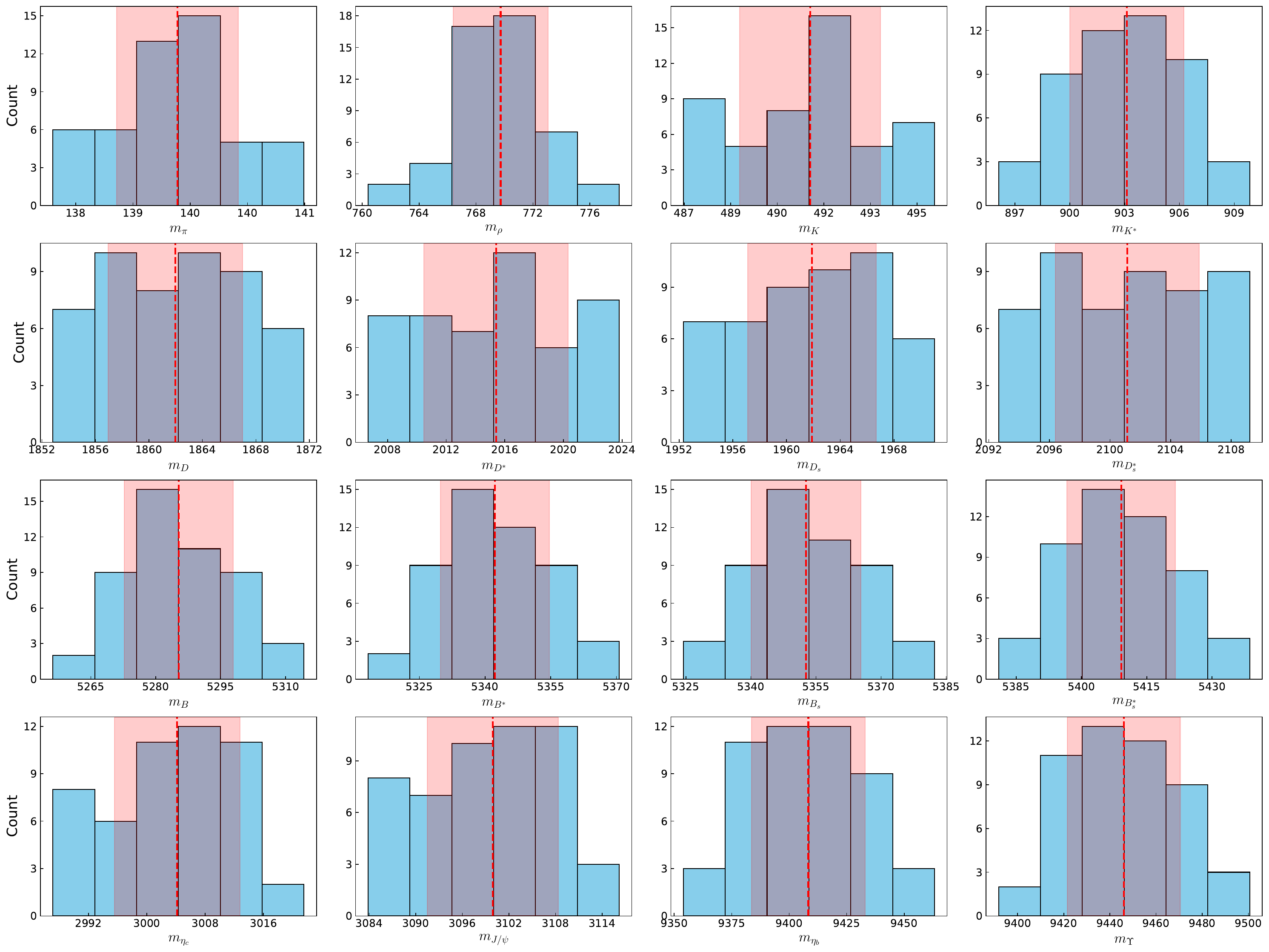}
	\caption{Distributions of meson mass spectra (in MeV) calculated using the 50 selected sets of AL1 parameters. The red dashed lines represent the mean values. The red shaded area corresponds to $\pm 1$ standard deviation.}
	\label{figsm:meson_dis}
\end{figure}

To be more specific, starting from the original parameter set, we generate new parameter sets by sampling each parameter independently from a Gaussian distribution centered at its original value, with a standard deviation equal to 0.5\% of that value. For each generated parameter set, we compute the corresponding mass spectra of sixteen ground-state mesons, including $\pi, \rho, K,K^*,D,D^*,D_s,D_s^*,B,B^*,B_s,B_s^*,\eta_c,J/\psi,\eta_b,\Upsilon$, and calculate the $\chi^2$ with respect to the experimental data,
\begin{equation}
	\chi^2=\sum_{i=1}^{16}\left(\frac{m_i^{\text{Theo.}}-m_i^{\text{Exp.}}}{\Delta m_i}\right)^2,
\end{equation}
where $\Delta m_i$ is taken to be $0.01 m_i^{\text{Exp.}}$ to account for the model uncertainty. We then compare the $\chi^2$ of the new parameter sets with the original one, denoted by $\chi_0^2$. A new parameter set is accepted if it satisfies the criterion,
\begin{equation}
	\frac{\chi^2-\chi_0^2}{\chi^2}<0.2,
\end{equation}
i.e., the quality of the description of the ground-state meson spectra remains comparable to that of the original parameter set. Applying this selection criterion, we retain 50 sets of parameters out of around 1500 generated sets. The retained parameter sets are taken to be the variation range of the parameters and are used to estimate the uncertainties. The distribution of the parameters and the meson spectra are shown in Figs.~\ref{figsm:para_dis} and \ref{figsm:meson_dis}.

\begin{figure}[ht]
	\centering
	\includegraphics[width=1\textwidth]{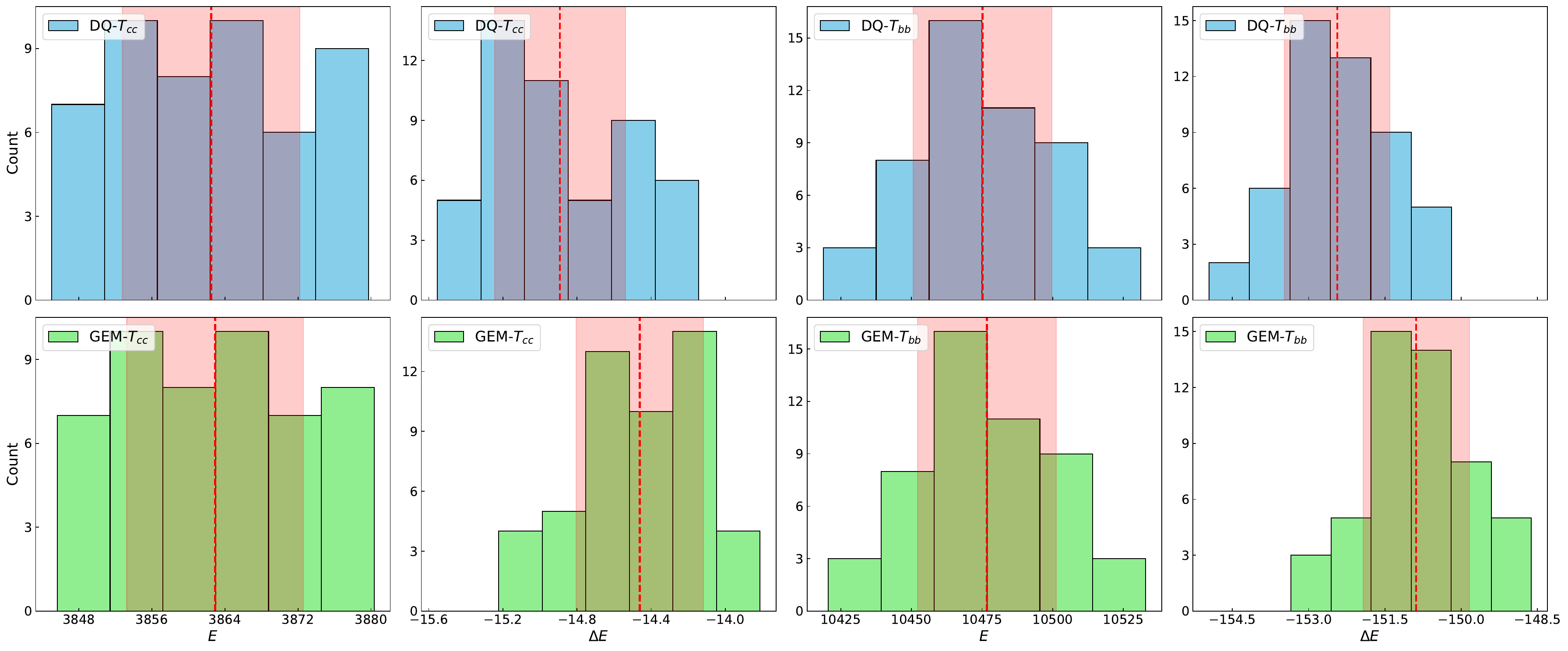}
	\caption{Distributions of ground-state energies $E$ and binding energies $\Delta E$ (in MeV) of the $I(S^P)=0(1^+)$ doubly heavy tetraquark bound states, calculated using the 50 selected sets of AL1 parameters. The first and second rows show the results from the DeepQuark (DQ) and Gaussian Expansion Method (GEM), respectively. The red dashed lines represent the mean values. The red shaded area corresponds to $\pm 1$ standard deviation. }
	\label{figsm:tetra_var}
\end{figure}

\begin{figure}[ht]
	\centering
	\includegraphics[width=1\textwidth]{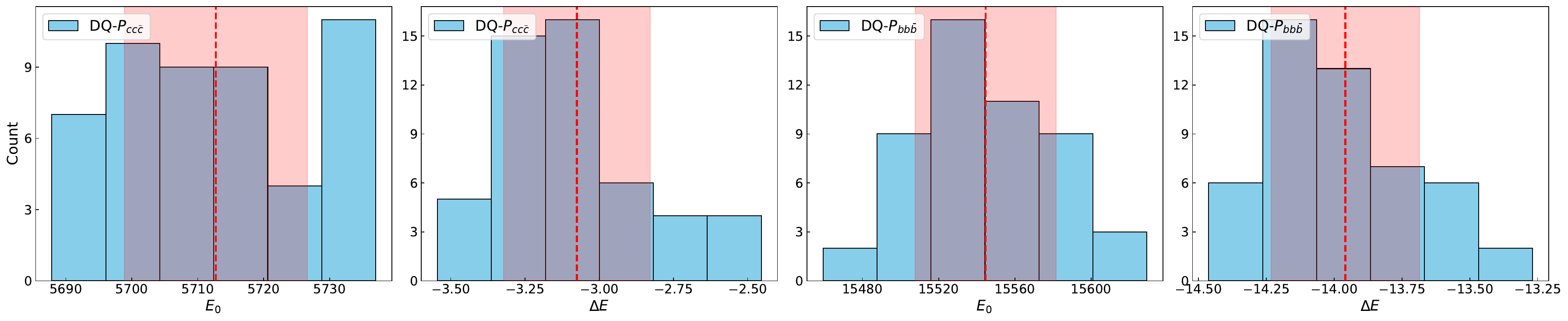}
	\caption{Distributions of ground-state energies $E$ and binding energies $\Delta E$ (in MeV) of the $I(S^P)=0(\frac{5}{2}^-)$ triply heavy pentaquark bound states, calculated using the 50 selected sets of AL1 parameters. The red dashed lines represent the mean values. The red shaded area corresponds to $\pm 1$ standard deviation.}
	\label{figsm:penta_var}
\end{figure}

\begin{table}[h]
	\centering
	\caption{The mean values and standard deviations of multiquark ground-state energies $E$ and binding energies $\Delta E$ calculated using the 50 selected sets of AL1 parameters. For $I(S^P)=0(1^+)$ doubly heavy tetraquark systems $T_{cc}$ and $T_{bb}$, we compare the results from the Gaussian Expansion Method (GEM) and DeepQuark (DQ). For $I(S^P)=0(\frac{5}{2}^-)$ triply heavy pentaquark systems $P_{cc\bar c}$ and $P_{bb\bar b}$, the DeepQuark framework is used to solve the ground states.}
	\begin{tabular}{cccc}
		\hline\hline
		Systems& Method&$E$ [MeV]&$\Delta E$ [MeV]\\
		\hline
		\multirow{2}{*}{$T_{cc}$}&GEM&$3863\pm10$&$-14.5\pm0.3$  \\
		&DQ&$3862\pm10$&$-14.9\pm0.3$\\
		\multirow{2}{*}{$T_{bb}$}&GEM&$10477\pm24$&$-150.9\pm1.0$ \\
		&DQ&$10475\pm24$&$-152.4\pm1.0$ \\
		$P_{cc\bar c}$&DQ&$5713\pm14$&$-3.1\pm0.2$\\
		$P_{bb\bar b}$&DQ&$15544\pm37$&$-13.9\pm0.3$\\
		\hline\hline
	\end{tabular}
	\label{tabsm:para_var}
\end{table}
We further calculate the multiquark bound states using the 50 selected parameter sets. We calculate the isoscalar vector doubly heavy tetraquark systems $T_{cc}$ and $T_{bb}$ using the GEM and the DeepQuark framework, whose results are shown in Fig.~\ref{figsm:tetra_var}. We also calculate ground-state energies of the triply heavy pentaquark bound states $P_{cc\bar c}$ and $P_{bb\bar b}$ using the DeepQuark framework, whose results are shown in Fig.~\ref{figsm:penta_var}. In Table~\ref{tabsm:para_var} we summarize the obtained results, which demonstrate the robustness of our multiquark bound state results and the DeepQuark framework.

\begin{itemize}
	\item Within the variation range of the model parameters, the energies of multiquark bound states may vary up to tens of MeV. However, the binding energies $\Delta E$ remain rather stable. The standard deviations of binding energies are around 1 MeV for deeply bound $T_{bb}$ state, and only around 0.3 MeV for shallow bound states $T_{cc}, P_{cc\bar c}, P_{bb\bar b}$. This shows the the strength of interactions and the existence of bound states are stable against the changes of parameters.
	\item Comparing the DeepQuark results with the GEM results in the doubly heavy tetraquark systems (the first and second row in Fig.~\ref{figsm:tetra_var}), we see that the DeepQuark results are in agreement with the GEM results for all 50 parameter sets. This shows that our neural-network framework can consistently describe the ground states of multiquark Hamiltonians with different parameter sets. Moreover, the DeepQuark energy result is slightly lower than the GEM energy result under the same parameter set. The mean values of binding energies is 0.4 MeV lower in the $T_{cc}$ system and 1.5 MeV lower in $T_{bb}$ system, as shown in Table~\ref{tabsm:para_var}. This is consistent with the our finding in the manuscript, demonstrating the strong expressive power of DeepQuark.  
\end{itemize}
\newpage
\clearpage

\subsection{Applications to other models}~\label{app:othermodel}
\begin{table}[h]
	\centering
	\caption{Different ingredients in five different quark models.}
	\begin{tabular}{|c|c|c|c|c|c|}
		\hline 
		& \multicolumn{2}{c|}{One-gluon-exchange} & \multicolumn{2}{c|}{Confinement} & \multirow{2}{*}{meson-exchange}\tabularnewline
		\cline{1-5}
		& asymptotic freedom & Hyperfine & form & screening effect & \tabularnewline
		\hline 
		\hline 
		AL1 & No & Gaussian & linear & No & No\tabularnewline
		\hline 
		AL2 & Yes & Gaussian & linear & No & No\tabularnewline
		\hline 
		AP1 & No & Gaussian & $p=2/3$ & No & No\tabularnewline
		\hline 
		AP2 & Yes & Gaussian & $p=2/3$ & No & No\tabularnewline
		\hline 
		SLM & No & Yukawa & linear & Yes & Yes\tabularnewline
		\hline 
	\end{tabular}
	\label{tabsm:5model}
\end{table}

In this work, we present the DeepQuark framework and employ the AL1 model to study the multiquark bound states. However, it should be noted that DeepQuark is not restricted to a specific model, but can be applied to investigate and discriminate among a variety of potential models with different ingredients. To demonstrate this, we employ DeepQuark to calculate the isoscalar vector doubly bottomed tetraquark $T_{bb}$ in five different quark potential models, including the AL1, AP1, AL2, AP2 models introduced in Ref.~\cite{SilvestreBrac1996} and the chiral quark model purposed by the group at Salamanca University in Refs.~\cite{Vijande:2004he,Segovia:2011dg} (denoted as SLM). The different ingredients included in the five models are listed in Table~\ref{tabsm:5model}.

The first four models feature one-gluon-exchange interaction and confinement interaction, taking the form
\begin{equation}
	\label{eqsm:pureQM}
	V_{ij} =-\frac{3}{16} \boldsymbol\lambda_i \cdot \boldsymbol\lambda_j\left(-\frac{\kappa(1-\exp(-r_{ij}/r_c))}{r_{i j}}+\frac{8 \pi \kappa^{\prime}(1-\exp(-r_{ij}/r_c))}{3 m_i m_j} \frac{e^{ -r_{i j}^2 / r_0^2}}{\pi^{3 / 2} r_0^3}\boldsymbol{s}_i \cdot \boldsymbol{s}_j-\Lambda+\lambda r_{i j}^p\right).
\end{equation}  
The four models differ by the confinement interaction and coupling coefficients. In the AL1 and AL2 models, the power of the confining term is  $p=1$, i.e. linear confinement is adopted. In the AP1 and AP2 models, $p=\frac{2}{3}$-power confining term is taken to give the correct asymptotic Regge trajectories. On the other hand, the AL1/AP1 model differs from the AL2/AP2 model by the coupling coefficients of the one-gluon-exchange interaction. In the AL1 and AP1 model, $r_c=0$ is taken, namely the coupling coefficients of the Coulomb and hyperfine terms (the first and second terms in Eq.~\eqref{eqsm:pureQM}) are taken to be constants. In the AL2 and AP2 model, $r_c\neq0$ is introduced as a phenomenological simulation of the asymptotic freedom. 

The other category of the quark potential model includes one-gluon-exchange interaction, confinement interaction, and the additional one-boson-exchange (OBE) interactions. This is called the chiral quark model, owing to the inclusion of the pseudoscalar meson exchange interactions arising from the spontaneous breaking of chiral symmetry. We choose the SLM model as a representative, whose interactions read
\begin{equation}
	\label{eqsm:chQM}
	V_{ij}=\boldsymbol\lambda_i \cdot \boldsymbol\lambda_j\left[\frac{\alpha_s}{4}\left(\frac{1}{r_{ij}}-\frac{2}{3m_im_j}\frac{e^{-r_{ij}/r_0}}{r_0^2r_{ij}}\boldsymbol{s}_i \cdot \boldsymbol{s}_j\right)+(-a_c(1-e^{-\mu_cr_{ij}})+\Delta)+V_{ij}^{\text{OBE}}\right].
\end{equation}\\
The regularization of the hyperfine term is taken to be the Yukawa type instead of the Gaussian type in Eq.~\eqref{eqsm:pureQM}. In addition, a confinement interaction with color screening effects is adopted, which becomes a linear interaction at short distance and a constant at long distance. The OBE interactions include the pseudoscalar meson exchange stemming from the spontaneous breaking of the chiral symmetry, and the scalar meson exchange, which is used to mimic the two-pion-exchange interaction. 
\begin{equation}
	\label{eqsm:OBE}
	V_{i j}^{\text{OBE}}=  V_{i j}^\pi \sum_{a=1}^3\left(\lambda_i^{f,a} \cdot \lambda_j^{f,a}\right)+V_{i j}^K \sum_{a=4}^7\left(\lambda_i^{f,a} \cdot \lambda_j^{f,a}\right) +V_{i j}^\eta\left[\cos \theta_P\left(\lambda_i^{f,8} \cdot \lambda_j^{f,8}\right)-\sin \theta_P\right] +V_{i j}^\sigma,
\end{equation}
with
\begin{equation}
	\begin{aligned}
		V_{i j}^\chi= & \frac{g_{\mathrm{ch}}^2}{4 \pi} \frac{m_\chi^2}{12 m_i m_j} \frac{\Lambda_\chi^2}{\Lambda_\chi^2-m_\chi^2} m_\chi\left(\boldsymbol{\sigma}_i \cdot \boldsymbol{\sigma}_j\right)\left[Y\left(m_\chi r_{i j}\right)-\frac{\Lambda_\chi^3}{m_\chi^3} Y\left(\Lambda_\chi r_{i j}\right)\right],\quad\chi=\pi, K, \eta \\
		V_{i j}^\sigma= & -\frac{g_{\mathrm{ch}}^2}{4 \pi} \frac{\Lambda_\sigma^2}{\Lambda_\sigma^2-m_\sigma^2} m_\sigma\left[Y\left(m_\sigma r_{i j}\right)-\frac{\Lambda_\sigma}{m_\sigma} Y\left(\Lambda_\sigma r_{i j}\right)\right],
	\end{aligned}
\end{equation}
where $\lambda^{f,a}$ is the Gell-Mann matrix in SU(3) flavor symmetry and $Y(x)=e^{-x}/x$ is the Yukawa function. The model parameters are taken from Ref.~\cite{Segovia:2011dg}.

\begin{table}[h]
	\centering
	\caption{Ground-state energies (in MeV) of the $I(S^P)=0(1^+)$ doubly bottomed tetraquark systems in various quark potential models. The binding energies $\Delta E$ are with respect to the lowest dimeson thresholds $(\bar B^*\bar B)$. The ground-state energies obtained from GEM~\cite{Meng:2023jqk} are listed for comparison.}
	\label{tabsm:Tbb}
	\begin{tabular}{lcccc}
		\hline\hline
		Potential&\multicolumn{2}{c}{DQ}&\multicolumn{2}{c}{GEM}~\cite{Meng:2023jqk}\\
		\hline
		&$E$&$\Delta E$&$E$&$\Delta E$\\
		\hline
		AL1&10491&-153&10493&-152\\
		AP1&10502&-176&10504&-175\\
		AL2&10473&-151& 10474&-151\\
		AP2&10471&-174& 10472&-174\\ 
		SLM&10230&-362&10232&-360\\
		\hline\hline
	\end{tabular}
\end{table}

We calcualte the ground-state energies of $T_{bb}$ in the five models, and compare the DeepQuark results with the GEM results in Table~\ref{tabsm:Tbb}. We find that DeepQuark can be successfully applied to all the models considered above, yielding ground-state energies that are consistent with and slightly lower than the corresponding GEM results. This demonstrates that DeepQuark can accommodate a wide variety of interactions, including different confining mechanisms and short-range dynamics, without being tied to a specific model. Combined with its capability to handle complex multiquark systems such as pentaquark states, DeepQuark provides a general and flexible framework for phenomenological studies of multiquark spectrum.

\section{Details of the optimization process}~\label{app:opt}

The deep neural network (DNN) in the DeepQuark wave function consists of the input features $\boldsymbol{x}=\left(\boldsymbol{r}_i,\left|\boldsymbol{r}_i-\boldsymbol{r}_j\right|,\alpha_c,\alpha_s,\alpha_t\right)$, four hidden fully connected layers and a one-dimensional output $f_{NN}(\boldsymbol{x})$. Here $\boldsymbol{r}_i,\alpha_c,\alpha_s,\alpha_t$ are the spatial three-dimensional coordinates in the center of mass frame and the encoded vectors of color, spin and isospin coupled bases, respectively. The number of nodes in each hidden layer and the total number of variational parameters in the DNN for different systems are given in Table~\ref{tabsm:DNN}. We use larger DNNs for larger systems to ensure the flexibility of the wave function.

\begin{table}[h]
	\centering
	\caption{The number of nodes in each hidden layer and the total number of variational parameters in the DNN for different systems.}
	\label{tabsm:DNN}
	\begin{tabular*}{\hsize}{@{}@{\extracolsep{\fill}}lccc@{}}
		\hline\hline
		Systems&$S^P$ &Nodes&Parameters \\
		\hline
		$ e^+e^- $&$0^+$&$(16,16,16,16)$&961\\
		$e^+e^-e^-$&$0^+$&$(16,16,16,16)$&1041\\
		$e^+e^+e^-e^-$&$0^+$ &$(16,16,16,16)$&1137\\
		$qqq$&$\frac{1}{2}^+$&$(16,16,16,16)$&1105\\
		$QQ\bar q\bar q$&$1^+$&$(32,16,16,16)$&1889\\
		$QQ\bar Q\bar Q$&$0^+$&$(32,16,16,16)$&1825\\
		$QQ\bar Q\bar Q$&$1^+$&$(32,16,16,16)$&1857\\
		$QQ\bar Q\bar Q$&$2^+$&$(32,16,16,16)$&1793\\
		$QQqq\bar Q$&$\frac{1}{2}^-$&$(40,20,20,20)$&3081\\
		$QQqq\bar Q$&$\frac{3}{2}^-$&$(40,20,20,20)$&3041\\
		$QQqq\bar Q$&$\frac{5}{2}^-$&$(40,20,20,20)$&2921\\
		\hline\hline
	\end{tabular*}
\end{table}

We optimize the variational parameters $\boldsymbol{\theta}$ by minimizing the energy expectation value. In each iteration step during the optimization process, we use the Metropolis-Hastings Monte Carlo algorithm~\cite{Metropolis1953,Hastings1970} to generate sample points that are distributed according to the probability
\begin{equation}
	P(R, \alpha)=\frac{\left|\psi_{\boldsymbol{\theta}}(R, \alpha)\right|^2}{\sum_\alpha \int d R\left|\psi_{\boldsymbol{\theta}}(R, \alpha)\right|^2},
\end{equation}
where $\psi_{\boldsymbol{\theta}}$ is the antisymmetric wave function,  $R=(\boldsymbol{r}_1,\cdots,\boldsymbol{r}_N)$ and $\alpha=(\alpha_c,\alpha_s,\alpha_t)$ denote the continuous and discrete degrees of freedom, respectively. Each Metropolis step consists of a Gaussian kick for the spatial coordinates and a random flip of the coupled channels $\alpha$~\cite{Adams:2020aax}. The newly proposed configuration $(R^\prime,\alpha^\prime)$ is accepted with probabilities
\begin{equation}
	P=\frac{\left|\psi_{\boldsymbol{\theta}}\left(R^{\prime}, \alpha^{\prime}\right)\right|^2}{\left|\psi_{\boldsymbol{\theta}}(R, \alpha)\right|^2}.
\end{equation}

The energy expectation value $E_{\boldsymbol{\theta}}=\frac{\langle\psi_{\boldsymbol{\theta}}|H|\psi_{\boldsymbol{\theta}}\rangle}{\langle\psi_{\boldsymbol{\theta}}|\psi_{\boldsymbol{\theta}}\rangle}$ and its gradient with respect to the parameters $\nabla_{\boldsymbol{\theta}} E_{\boldsymbol{\theta}}$ can be estimated using these sample points.
\begin{equation}
	\begin{aligned}
		E_{\boldsymbol{\theta}}&=\frac{\sum_{\alpha\alpha^\prime}\int dR\, \psi_{\boldsymbol{\theta}}^*(R,\alpha)H_{\alpha\alpha^\prime}(R)\psi_{\boldsymbol{\theta}}(R,\alpha^\prime)}{\sum_\alpha\int dR\,|\psi_{\boldsymbol{\theta}}(R,\alpha)|^2}\\
		&=\sum_{\alpha\alpha^\prime}\int dR\,P(R,\alpha)\psi_{\boldsymbol{\theta}}^{-1}(R,\alpha)H_{\alpha\alpha^\prime}(R)\psi_{\boldsymbol{\theta}}(R,\alpha^\prime)\\
		&=\frac{1}{N}\sum_n E_L(R_n,\alpha_n),
	\end{aligned}
\end{equation}
where $N$ is the number of sample points $(R_n,\alpha_n)$, and the local energy $E_L$ is given by,
\begin{equation}
	E_L(R,\alpha)=\sum_{\alpha^\prime}\psi_{\boldsymbol{\theta}}(R,\alpha)^{-1}H_{\alpha\alpha^\prime}(R)\psi_{\boldsymbol{\theta}}(R,\alpha^\prime).
\end{equation}
Similarly, the gradient can be estimated as,
\begin{equation}
	\begin{aligned}
		\nabla_{\boldsymbol{\theta}} E_{\boldsymbol{\theta}}=2 &\left[\frac{1}{N}\left(\sum_n E_L(R_n,\alpha_n)\nabla_{\boldsymbol{\theta}}\log\psi_{\boldsymbol{\theta}}(R_n,\alpha_n)\right)\right.-\\
		&\left.\,\frac{1}{N^2}\left(\sum_n E_L(R_n,\alpha_n)\right)\left(\sum_n\nabla_{\boldsymbol{\theta}}\log\psi_{\boldsymbol{\theta}}(R_n,\alpha_n)\right)\right].
	\end{aligned}
\end{equation}
The parameters are then updated using the stochastic reconfiguration~\cite{PhysRevB.71.241103}, a commonly used optimization method in VMC~\cite{Adams:2020aax,Yang:2022rlw},
\begin{equation}
	\boldsymbol{\theta}^{i+1}=\boldsymbol{\theta}^i-\eta(S+\epsilon I)^{-1}\nabla_{\boldsymbol{\theta}^i}E_{\boldsymbol{\theta}^i},
\end{equation}
where $i$ is the iteration step, $\eta$ is the learning rate, $\epsilon=10^{-2}$ is taken for numerical stability, and $S$ is the Quantum Fisher information matrix,
\begin{equation}
	S_{ab}=\frac{\langle \partial_{\theta_a}\psi_{\boldsymbol{\theta}}|\partial_{\theta_b}\psi_{\boldsymbol{\theta}}\rangle}{\langle\psi_{\boldsymbol{\theta}}|\psi_{\boldsymbol{\theta}}\rangle}-\frac{\langle \partial_{\theta_a}\psi_{\boldsymbol{\theta}}|\psi_{\boldsymbol{\theta}}\rangle}{\langle\psi_{\boldsymbol{\theta}}|\psi_{\boldsymbol{\theta}}\rangle}\frac{\langle \psi_{\boldsymbol{\theta}}|\partial_{\theta_b}\psi_{\boldsymbol{\theta}}\rangle}{\langle\psi_{\boldsymbol{\theta}}|\psi_{\boldsymbol{\theta}}\rangle}.
\end{equation}
The NetKet package~\cite{netket3:2022} provides efficient implementation of the stochastic reconfiguration algorithm.

During initial optimization, the trial wave function resides far from the ground state in parameter space. We generate a small sample of $N=2\times10^4$ points to rapidly estimate $E_{\boldsymbol{\theta}}$ and $\nabla_{\boldsymbol{\theta}} E_{\boldsymbol{\theta}}$ with deliberate coarseness, enabling efficient early-stage convergence. As the DeepQuark wave function approaches convergence, we increase the sample size to stabilize the optimization process. During the final optimization stage, we save 10 distinct sets of wave function parameters at a 50-step interval.   The energy and other physical observables of these wave function configurations are subsequently evaluated using $N\sim10^6$ sample points. The parameter set yielding the lowest energy expectation value is selected as the final ground-state wave function.

For each multiquark system, it only take merely $\mathcal{O}(1)$ A100 GPU-hours to get ground state solution. Although direct comparison between phenomenological models and ab initio simulations is improper, we provide lattice QCD ensemble generation costs solely to contextualize computational scales. For example, it was reported  the CLQCD collaboration spent over $6\times 10^5$ NVIDIA A100 GPU-hours to generate only part of the  planned ensembles~\cite{YiBoYangQuarkMassLow2023}.

\section{Results of electron systems}~\label{app:eee}
The ground-state energy results of few-electron systems are listed in Table~\ref{tabsm:leptons}. 
For comparative purposes, we include well-established benchmark results such as the exact solution of $\mathrm{Ps}$, and high-precision variational results for $\mathrm{Ps}^-$ using  Hylleraas-type wave functions~\cite{Ho:1993zz} and $\mathrm{Ps}_2$ using explicitly correlated Gaussian functions~\cite{Kinghorn:1993zz,PhysRevA.74.052502}.

\begin{table}[h]
	\centering
	\caption{Ground-state energies (in eV) of the few-electron systems. The statistical errors of the DeepQuark results are shown in the parentheses. Results from other methods are shown for comparison. $\mathrm{Ps}$ can be solved exactly while $\mathrm{Ps}^-$ and $\mathrm{Ps}_2$ are solved using variational method.}
	\label{tabsm:leptons}
	\begin{tabular*}{\hsize}{@{}@{\extracolsep{\fill}}lcc@{}}
		\hline\hline
		& DQ&Other Methods \\
		\hline
		$ e^+e^- $&-6.80301(16)&-6.803\footnote{exact}\\
		$e^+e^-e^-$&-7.12882(16)&-7.130~\cite{Ho:1993zz}\\
		$e^+e^+e^-e^-$ &-14.0347(7)&-14.04~\cite{Kinghorn:1993zz,PhysRevA.74.052502}\\
		\hline\hline
	\end{tabular*}
\end{table}

\end{document}